\documentclass[prb,twocolumn,amsmath,amssymb,superscriptaddress]{revtex4-1}
\renewcommand{\figurename}{Fig.\!}

\usepackage[normalem]{ulem}
\usepackage{graphicx}
\usepackage{todonotes} 
\usepackage{dcolumn}
\usepackage{tabularx}
\usepackage{bm}
\usepackage{hyperref}
\usepackage{textcomp}
\hypersetup{
    colorlinks=true,
    linkcolor=blue,
    citecolor = blue,
    filecolor=magenta,      
    urlcolor=blue,
    pdftitle={Programmable Heisenberg interactions between Floquet qubits},
    pdfpagemode=FullScreen,
    }

\usepackage{multirow}
\usepackage{nicefrac}
\usepackage{qcircuit}
\usepackage{physics}

\begin{document}


\title{Programmable Heisenberg interactions between Floquet qubits}

\author{Long B. Nguyen}
\email{nbaolong89@gmail.com}
\thanks{Equal contribution.}
\affiliation{Computational Research Division, Lawrence Berkeley National Laboratory, Berkeley, California 94720, USA}
\affiliation{Department of Physics, University of California, Berkeley, California 94720, USA}

\author{Yosep Kim}
\email{yosep9201@gmail.com}
\thanks{Equal contribution.}
\affiliation{Computational Research Division, Lawrence Berkeley National Laboratory, Berkeley, California 94720, USA}
\affiliation{Department of Physics, University of California, Berkeley, California 94720, USA}
\affiliation{Center for Quantum Information, Korea Institute of Science and Technology (KIST), Seoul 02792, Korea}

\author{Akel Hashim}
\affiliation{Computational Research Division, Lawrence Berkeley National Laboratory, Berkeley, California 94720, USA}
\affiliation{Department of Physics, University of California, Berkeley, California 94720, USA}

\author{Noah Goss}
\affiliation{Computational Research Division, Lawrence Berkeley National Laboratory, Berkeley, California 94720, USA}
\affiliation{Department of Physics, University of California, Berkeley, California 94720, USA}

\author{Brian Marinelli}
\affiliation{Computational Research Division, Lawrence Berkeley National Laboratory, Berkeley, California 94720, USA}
\affiliation{Department of Physics, University of California, Berkeley, California 94720, USA}

\author{\\ Bibek Bhandari}
\affiliation{Institute for Quantum Studies, Chapman University, Orange, CA 92866, USA}

\author{Debmalya Das}
\affiliation{Department of Physics and Astronomy, University of Rochester, Rochester, NY 14627, USA}
\affiliation{Institute for Quantum Studies, Chapman University, Orange, CA 92866, USA}


\author{Ravi K. Naik}
\affiliation{Computational Research Division, Lawrence Berkeley National Laboratory, Berkeley, California 94720, USA}
\affiliation{Department of Physics, University of California, Berkeley, California 94720, USA}

\author{John Mark Kreikebaum}
\affiliation{Department of Physics, University of California, Berkeley, California 94720, USA}
\affiliation{Materials Science Division, Lawrence Berkeley National Laboratory, Berkeley, California 94720, USA}


\author{\\Andrew N. Jordan}
\affiliation{Institute for Quantum Studies, Chapman University, Orange, CA 92866, USA}
\affiliation{Department of Physics and Astronomy, University of Rochester, Rochester, NY 14627, USA}

\author{David I. Santiago}
\affiliation{Computational Research Division, Lawrence Berkeley National Laboratory, Berkeley, California 94720, USA}
\affiliation{Department of Physics, University of California, Berkeley, California 94720, USA}

\author{Irfan Siddiqi}
\affiliation{Computational Research Division, Lawrence Berkeley National Laboratory, Berkeley, California 94720, USA}
\affiliation{Department of Physics, University of California, Berkeley, California 94720, USA}
\affiliation{Materials Science Division, Lawrence Berkeley National Laboratory, Berkeley, California 94720, USA}


\begin{abstract}
    
The fundamental trade-off between robustness and tunability is a central challenge in the pursuit of quantum simulation and fault-tolerant quantum computation. In particular, many emerging quantum architectures~\cite{gyenis2021moving,siddiqi2021engineering} are designed to achieve high coherence at the expense of having fixed spectra and consequently limited types of controllable interactions. 
Here, by adiabatically transforming fixed-frequency superconducting circuits into modifiable Floquet qubits, we demonstrate an $\mathrm{XXZ}$ Heisenberg interaction with fully adjustable anisotropy. This interaction model is on one hand the basis for many-body quantum simulation of spin systems~\cite{savary2016quantum}, and on the other hand the primitive for an expressive quantum gate set~\cite{kivlichan2018quantum}.
To illustrate the robustness and versatility of our Floquet protocol, we tailor the Heisenberg Hamiltonian and implement two-qubit $\mathrm{iSWAP}$, $\mathrm{CZ}$, and $\mathrm{SWAP}$ gates with estimated fidelities of $99.32(3)\%$, $99.72(2)\%$, and $98.93(5)\%$, respectively. In addition, we implement a Heisenberg interaction between higher energy levels and employ it to construct a three-qubit $\mathrm{CCZ}$ gate with a fidelity of $96.18(5)\%$. Importantly, the protocol is applicable to various fixed-frequency high-coherence platforms, thereby unlocking a suite of essential interactions for high-performance quantum information processing. From a broader perspective, our work provides compelling avenues for future exploration of quantum electrodynamics and optimal control using the Floquet framework~\cite{shirley1965solution}.

\end{abstract}

\maketitle


\vspace{1em}
\noindent\textbf{Introduction}

\noindent The capability to coherently choreograph interactions between qubits is the foundation for the recent advances in quantum technologies. A quintessential example is the manipulation of the quantum Heisenberg model for the simulation of many-body quantum spin systems~\cite{savary2016quantum,bernien2017probing,jepsen2020spin}, which has led to the recent discoveries of intriguing physical phenomena such as discrete time crystal~\cite{zhang2017observation}, phantom spin-helix states~\cite{jepsen2022long}, and formation of photon bound states~\cite{morvan2022formation}. The Heisenberg interactions are also the primitives for expressive multi-qubit gates~\cite{kivlichan2018quantum} which play important roles in quantum algorithms~\cite{bharti2022noisy} and quantum error correction~\cite{fowler2012surface,ghosh2015leakage}. Therefore, endowing quantum architectures with such archetypal interactions significantly extends their capabilities and performance.



The required tunability in solid-state quantum devices generally entails additional decoherence channels, demanding design overhead, and increased operational complexity. For example, in the domain of superconducting circuits, the performance in flux-tunable devices is typically limited by unavoidable $1/f$ noise arising from the surrounding environment. Meanwhile, fixed-frequency platforms such as single-junction transmon~\cite{koch2007charge,paik2011observation} and fluxonium~\cite{manucharyan2009fluxonium} biased at the half-integer flux quantum~\cite{nguyen2019high} have the best coherence times to date, but their native interactions are limited to the cross-resonance~\cite{chow2011simple,kim2022high} and longitudinal couplings~\cite{mitchell2021hardware,wei2022hamiltonian}. Introducing additional tunable couplers enables a parametric transverse coupling, but the performance is undermined by the couplers' coherence and spurious couplings~\cite{ganzhorn2020benchmarking}.


 In this work, we present a reliable and hardware-efficient protocol to synthesize Floquet qubits~\cite{huang2021engineering,gandon2022engineering} from statically coupled single-junction transmon qubits using time-periodic microwave drives, showing that the adiabatic mapping procedure can be hastened by exploiting a shortcuts-to-adiabaticity (STA) technique~\cite{gu2019shortcuts,motzoi2009simple,gambetta2011analytic}. Then, we implement an $\mathrm{XXZ}$ Heisenberg interaction between these Floquet qubits, described by the Hamiltonian
\begin{equation}
    \hat{\mathcal{H}}_\mathrm{XXZ}/\hbar =\sum_{ i,j} J_\mathrm{XY}(\hat{\sigma}_\text{x}^{i}\hat{\sigma}_\text{x}^{j} + \hat{\sigma}_\text{y}^{i}\hat{\sigma}_\text{y}^{j}) + J_\mathrm{ZZ}\hat{\sigma}_\text{z}^{i}\hat{\sigma}_\text{z}^{j},
    \label{eqn:xxz}
\end{equation}
and demonstrate that the transverse spin-exchange and longitudinal spin-spin interaction terms can be adjusted independently by tailoring the drive parameters.

To validate the robustness and practicality of the protocol, we characterize two-qubit $\mathrm{iSWAP}$, $\mathrm{CZ}$, and $\mathrm{SWAP}$ gates which correspond to different anisotropy $J_\mathrm{ZZ}/J_\mathrm{XY}$, achieving estimated fidelities of $99.32(3)\%$, $99.72(2)\%$, and $98.93(5)\%$, respectively. In addition, we show that the Floquet-engineered interactions can be broadly applied to other levels in the system. Specifically, we explore the swapping between the qutrit states $|11\rangle$ and $|02\rangle$, then employ it to implement a three-qubit $\mathrm{CCZ}$ gate which is locally equivalent to the Toffoli gate~\cite{fedorov2011implementation}, achieving an estimated fidelity of $96.18(5)\%$. Our work exemplifies the operational principles of Floquet qubits and illustrates their broad potential, thus opening promising pathways for future developments of the Floquet framework in enhancing the capabilities of fixed-frequency solid-state quantum platforms.

\begin{figure*}[t]
    \includegraphics[width=\textwidth]{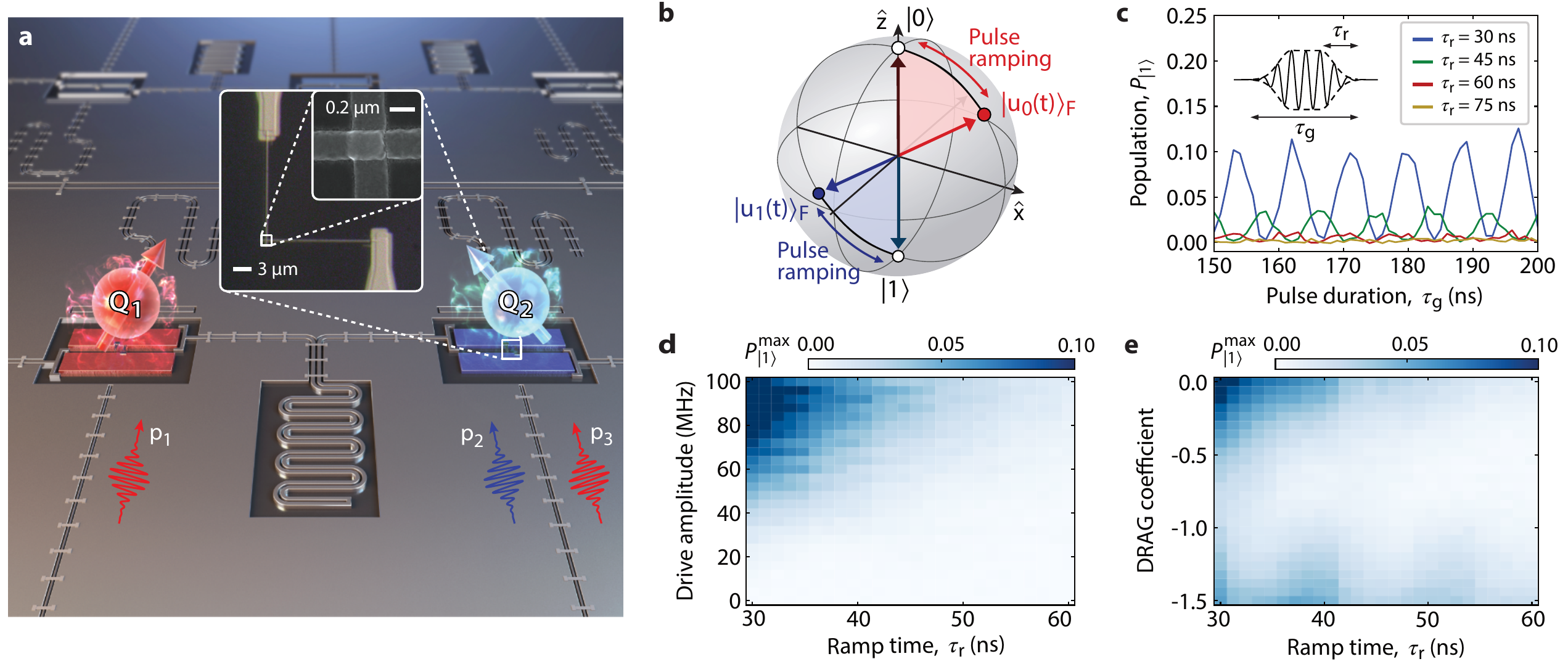}
    \caption{\label{fig1}\textbf{Floquet qubit}. 
    \textbf{a,} Experimental schematic depicting two single-junction transmon qubits $\text{Q}_1$ (red) and $\text{Q}_2$ (blue) coupled via a shared coplanar-waveguide resonator resulting in an effective static coupling~\cite{majer2007coupling}. Microwave pulses are applied to the transmission lines situated below the capacitor pads to sculpt the Floquet qubits and control single-qubit rotations. The Heisenberg interactions are programmed by tailoring pulses p$_1$, p$_2$, and p$_3$.
    \textbf{b,} Bloch sphere representation of the adiabatic transformation from a bare qubit in the lab frame to a Floquet qubit in the rotating frame. 
    \textbf{c,} Population in the excited state $P_{|1\rangle}$ of $\text{Q}_1$ when subjected to a microwave pulse with amplitude $100~\mathrm{MHz}$, detuning $-40~\mathrm{MHz}$, ramp time $\tau_\text{r}$, and duration $\tau_\text{g}$. Nonadiabaticity manifests as finite oscillations for short ramp time $\tau_\text{r}$.
    \textbf{d,} Dependence of the maximum state leakage $P^{\text{max}}_{|1\rangle}$ on the pulse's amplitude and ramp time $\tau_\text{r}$. The drive is applied at the same frequency from \textbf{c}. 
    \textbf{e,} Dependence of $P^{\text{max}}_{|1\rangle}$ on the pulse's ramp time $\tau_\text{r}$ and DRAG coefficient. The drive amplitude and frequency are the same from \textbf{c}.}
\end{figure*}





 
 
\vspace{0.5em}
\noindent\textbf{Synthesizing Floquet qubits}

\begin{figure*}[t]
    \includegraphics[width=\textwidth]{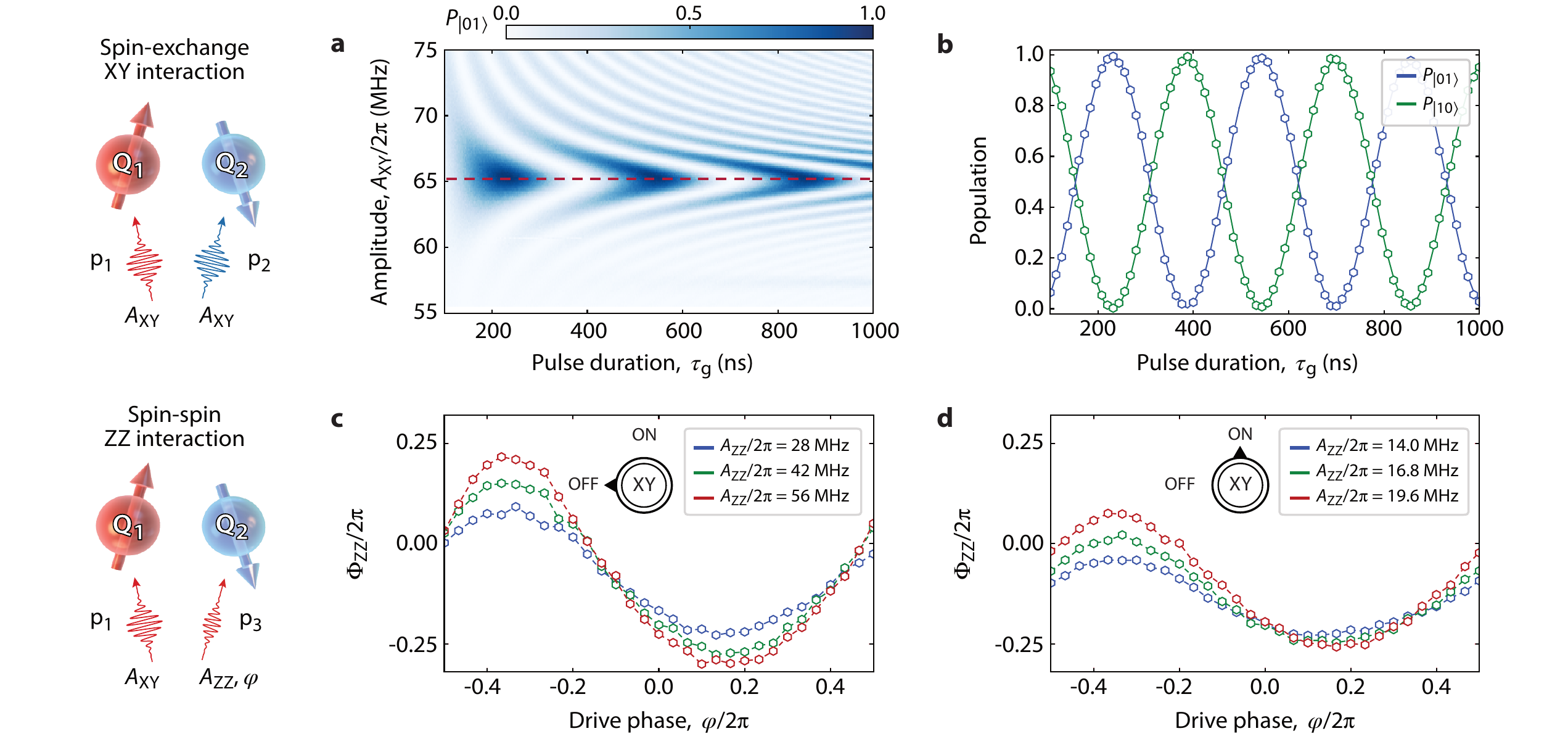}
    \caption{\label{fig2}\textbf{Floquet-engineered $\mathrm{XXZ}$ Heisenberg interaction}. \textbf{a,} Chevron pattern showing the dependence of $P_{|01\rangle}$ on the amplitude $A_\mathrm{XY}$ and duration $\tau_\text{g}$ of pulses p$_1$ and p$_2$ for qubits initialized in $|10\rangle$ (top left schematic). The pulses are applied at frequencies 40 MHz red(blue)-detuned from Q$_{1(2)}$'s $|0\rangle\leftrightarrow|1\rangle$ transition with DRAG coefficients $\lambda_\mathrm{DRAG}=-(+)0.6$ and ramp time $\tau_\text{r}=50~\mathrm{ns}$.
    \textbf{b,} Coherent oscillation between $|10\rangle$ and $|01\rangle$ for the optimal pulse amplitude $A_\mathrm{XY}/2\pi=65.2~\mathrm{MHz}$ highlighted by the red dashed line in \textbf{a}. 
    \textbf{c,} Extracted longitudinal coupling angle $\Phi_\mathrm{ZZ}$ between the qubits after p$_1$ and p$_3$ are applied at 40 MHz red-detuned from Q$_1$'s transition frequency (bottom left schematic). They are 200-ns-long, with ramp time $\tau_r=50~\mathrm{ns}$ and DRAG coefficient $\lambda_\mathrm{DRAG}=-0.6$. p$_1$'s amplitude is fixed at $A_\mathrm{XY}/2\pi=65.2~\mathrm{MHz}$, while p$_2$'s amplitude $A_\mathrm{ZZ}$ and phase $\varphi$ are varied to tune the $\mathrm{ZZ}$ coupling rate.
    \textbf{d,} Extracted longitudinal coupling angle $\Phi_\mathrm{ZZ}$ between the qubits after being subjected to all three pulses. The amplitude $A_\mathrm{XY}$ of p$_{1}$ and p$_{2}$ is tuned to induce a full $|10\rangle \leftrightarrow |01\rangle$ swap, while the amplitude $A_\mathrm{ZZ}$ and phase $\varphi$ of p$_3$ are varied to tune the $\mathrm{ZZ}$ rate during the swap. }
\end{figure*}

\noindent Figure~\ref{fig1}\textbf{a} depicts the superconducting device used in the experiment. It consists of single-junction transmon qubits~\cite{koch2007charge} which are pairwise coupled via mutual coplanar waveguide resonators~\cite{majer2007coupling,blais2021circuit}. Details of the quantum device and experimental setup are presented in Methods. Although the frequencies of the qubits are fixed after fabrication, Floquet engineering has recently emerged as a powerful tool that allows the sculpting of effective Hamiltonians that are otherwise unavailable~\cite{weitenberg2021tailoring}, thus promising a new dimension to tune the system. Here, we synthesize Floquet qubits using detuned periodic microwave drives and tailor them to engineer the Hamiltonian given by Eq.~(\ref{eqn:xxz}).

The mapping is described by Floquet formalism as follows. The Hamiltonian of a two-level spin-half system subjected to a periodic driving field with amplitude $A$, frequency $\omega_\text{d}$, and phase $\varphi$ is given as
\begin{equation}
\hat{\mathcal{H}}_\text{q}(t)/\hbar=-\frac{\omega_\text{q}}{2}\hat{\sigma}_\text{z}+ A\ \!\text{cos}(\omega_\text{d}t+\varphi)\ \!\hat{\sigma}_\text{x},
\label{Floquet_H}
\end{equation}
where $\hbar\omega_\text{q}$ is the energy gap of the two-level system, and
$\hat{\sigma}_\text{z}$ and $\hat{\sigma}_\text{x}$ represent the Pauli operators. There exists no static eigenenergies and eigenstates of the system as solutions of the time-dependent Schrödinger equation $i\hbar\partial_t|\psi(t)\rangle=\hat{\mathcal{H}}_\text{q}(t)|\psi(t)\rangle$. However, due to the periodicity of  $\hat{\mathcal{H}}_\text{q}(t)$, the Schrödinger equation can be modified into the Floquet equation~\cite{shirley1965solution,deng2016dynamics}, $\left(\hat{\mathcal{H}}_\text{q}(t)-i\hbar\partial_t\right)|u_n(t)\rangle_\text{F}=\hbar\varepsilon_n|u_n(t)\rangle_\text{F}$, and static quasienergies $\hbar\varepsilon_n$ can be found for time-periodic Floquet states $|u_n(t)\rangle_\text{F}=|u_n(t+2\pi/\omega_d)\rangle_\text{F}$. Here the Floquet states are denoted with subscript F to distinguish them from the bare states in the lab frame. The Floquet and bare states are interconvertible following the relation
\begin{equation}
    |u_n(t)\rangle_\text{F}= e^{i\varepsilon_nt}|\psi_n(t)\rangle.
    \label{Floquet_Eq}
\end{equation}

Interestingly, $e^{ik\omega_\text{d} t}|u_{n}(t)\rangle_\text{F}$ with integer $k$ also satisfies the Floquet equation and has quasienergy $\hbar(\varepsilon_n+k\omega_\text{d})$, resulting in an infinite transition spectrum~\cite{deng2016dynamics},
$\varepsilon_{1}-\varepsilon_{0}=k \omega_\text{d}\pm\sqrt{ A^2+(\omega_\text{d}-\omega_\text{q})^2}$, where the plus(minus) sign corresponds to red(blue) detuned drive. In addition, the drive phase $\varphi$ acts as a time translation operator on the Hamiltonian in Eq.~(\ref{Floquet_H}), $|u_n(t)\rangle_\text{F}\rightarrow|u_n(t+\varphi/\omega_\text{d})\rangle_\text{F}$. These show how the Floquet states and their quasienergies depend on the drive parameters $A$, $\varphi$, and $\omega_\text{d}$, which we can use to tailor the driven systems (See Methods for detailed Floquet formalism).



To prepare a Floquet qubit with the desired properties, we have to continuously map the undriven qubit to the Floquet basis, as shown in Fig.~\ref{fig1}\textbf{b}. If the transformation is performed abruptly, there exists finite tunneling between the Floquet basis states, and the process becomes nonadiabatic. According to Adiabatic Theorem~\cite{deng2016dynamics}, the tunneling rate is proportional to $dA/dt$, that is, the target Floquet qubit corresponding to a larger drive amplitude must be transformed using a longer ramp time.

We experimentally explore this by irradiating qubit Q$_1$ with a cosine-ramp pulse with different pulse durations $\tau_\text{g}$ and ramp times $\tau_\text{r}$, then measuring its excited state populartion $P_{|1\rangle}$. The drive amplitude is set to be $100~\mathrm{MHz}$ in terms of on-resonant Rabi frequency, and the drive frequency is red-detuned from Q$_1$'s transition frequency by $40~\mathrm{MHz}$. Nonadiabatic effects then manifest as finite excited state populations after the pulse, which oscillates with respect to the pulse duration due to the dynamical phase accumulation of the Floquet qubit (Fig.~\ref{fig1}\textbf{c}). Evidently, shorter ramp times correspond to more severe nonadiabatic effects. In addition, the result in Fig.~\ref{fig1}\textbf{d} confirms that a larger drive amplitude requires a longer ramp time to satisfy the adiabatic condition.


Interestingly, we find that using an STA technique known as derivative removal by adiabatic gate (DRAG)~\cite{motzoi2009simple,gambetta2011analytic} helps reduce nonadiabatic effects substantially. As shown in Fig.~\ref{fig1}\textbf{e}, the excited state leakage corresponding to a short-ramp pulse can be suppressed by adding a quadrature component to the pulse with amplitude $A_\text{Q}=\lambda_\mathrm{DRAG}\times {dA(t)}/dt$. Case in point, $\tau_\text{r}$ can be reduced from 60 ns to 30 ns by employing a DRAG coefficient $\lambda_\text{DRAG}=-0.7$. This suggests that advanced optimal control techniques can be explored to further accelerate the mapping procedure.


\vspace{1em}

\noindent\textbf{Tailoring Heisenberg interactions}

\noindent Having established the general conditions for adiabatic mapping between undriven qubit states and Floquet states, we next orchestrate the microwave pulses to engineer the $\mathrm{XXZ}$ Heisenberg interaction in Eq.~(\ref{eqn:xxz}) between the Floquet qubits. The interaction Hamiltonian describing the coupling between Q$_1$ and Q$_2$ in Fig.~\ref{fig1}\textbf{a}  is
    $\hat{\mathcal{H}}_\text{int}/\hbar= J\hat{\sigma}^{(1)}_\text{x}\hat{\sigma}^{(2)}_\text{x}$,
where $J$ is the static coupling strength, the superscripts are qubit indices, and the Pauli operators are defined in the undriven basis. This interaction can be described by a Floquet Hamiltonian using the relation given by Eq.~(\ref{Floquet_Eq}),
\begin{equation}
\label{eqn:floquet_heisenberg}
\hat{\mathcal{H}}_\text{int,F}/\hbar= J \sum_{a,b,c,d } c^{(1)}_{ab}(t)c^{(2)}_{cd}(t) e^{i(\varepsilon^{(1)}_{ab}+\varepsilon^{(2)}_{cd})t}\hat{f}^{(1)}_{ab}(t)\hat{f}^{(2)}_{cd}(t),
\end{equation} 
where $\varepsilon^{(k)}_{ab} \equiv \varepsilon^{(k)}_{b}-\varepsilon^{(k)}_{a}$, $c^{(k)}_{ab}(t)=\langle \psi_a^{(k)}(t)|\hat{\sigma}^{(k)}_\text{x}|\psi_b^{(k)}(t)\rangle$, $\hat{f}^{(k)}_{ab}(t)={}|u^{(k)}_a(t)\rangle_\text{F}\langle u^{(k)}_b(t)|_\text{F}$ for qubit Q$_k$, and $a,b,c,d \in \{0,1\}$ for two qubits. The fast oscillation dynamics can be neglected by invoking the rotating wave approximation, leaving only the terms that follow energy conservation law, $\varepsilon^{(1)}_{ab}+\varepsilon^{(2)}_{cd}=0$ for  $abcd \in$ $\{0110$, $1001$, $0000$, $0011$, $1100$, $1111\}$. Inspecting the reduced Hamiltonian then gives us insight on the types of interactions present between the qubits.

On one hand, the terms satisfying $\varepsilon^{\scalebox{.6}{(}1\scalebox{.6}{)}}_{01} = \varepsilon^{\scalebox{.6}{(}2\scalebox{.6}{)}}_{01}$ correspond to the transverse XY spin-exchange interaction in Eq.~(\ref{eqn:xxz}) with $J_\text{XY}=J \langle c_{01}^{\scalebox{.6}{(}1\scalebox{.6}{)}}c_{10}^{\scalebox{.6}{(}2\scalebox{.6}{)}}\rangle_t = J\langle c_{10}^{\scalebox{.6}{(}1\scalebox{.6}{)}}c_{01}^{\scalebox{.6}{(}2\scalebox{.6}{)}}\rangle_t$, where $\langle ...\rangle_t$ denotes the time-average value. This process follows the conventional wisdom that an XY exchange-type interaction between two coupled spins occurs when they are brought into resonance with each other. On the other hand, the rest of the reduced Hamiltonian produces the longitudinal ZZ spin-spin coupling in Eq.~(\ref{eqn:xxz}), $J_\text{ZZ}=J\langle c_{11}^{(1)}c_{11}^{(2)}+c_{00}^{(1)}c_{00}^{(2)}-c_{00}^{(1)}c_{11}^{(2)}-c_{11}^{(1)}c_{00}^{(2)}\rangle_t$. Consequently, we can program the transverse and longitudinal interactions independently by tailoring the quasienergies with periodic microwave drives.

We proceed to validate this principle as follows. First, we engineer a pure transverse XY spin-exchange interaction corresponding to an XX Heisenberg model where the anisotropy is zero, $\Delta=J_\mathrm{ZZ}/J_\mathrm{XY}=0$. Given that Q$_1$'s frequency is lower than that of Q$_2$, their quasienergy differences $\varepsilon^{\scalebox{.6}{(}k\scalebox{.6}{)}}_{01}$ can be brought into resonance if Q$_1$(Q$_2$) is driven with red(blue) detuned microwaves (Extended Data Fig.~\ref{figsf}\textbf{b}). After preparing the qubits in $|10\rangle$, we apply two such pulses (p$_1$ and p$_2$ in Fig.~\ref{fig1}\textbf{a} and top left panel of Fig.~\ref{fig2}) with the same duration $\tau_\text{g}$ and amplitude $A_\mathrm{XY}$ at a detuning frequency of 40 MHz. 
We observe a coherent population transfer to state $|01\rangle$ that forms a chevron pattern as a function of $\tau_\text{g}$ and $A_\mathrm{XY}$, signifying a transverse coupling between the qubits (Fig.~\ref{fig2}\textbf{a}). Notably, although the interaction occurs between the Floquet qubits in the dressed frame, the adiabatic connection ascertains the exchange between the bare qubit states after the reverse mapping, which bears resemblance to the latching mechanism in classical electronics. Indeed, at the optimal drive amplitude $A_\mathrm{XY}/2\pi=65.2~\mathrm{MHz}$ (Fig.~\ref{fig2}\textbf{b}), $|10\rangle$ and $|01\rangle$ exhibit coherent oscillations at a rate of $3.2$~MHz, which is limited by the static coupling constant $J$ (Methods). The lack of fast oscillatory behavior is a clear indication of the high mapping fidelity.


Next, we proceed to induce a pure longitudinal $\mathrm{ZZ}$ spin-spin coupling corresponding to an Ising interaction between the Floquet qubits. This can be accomplished by irradiating microwave drives p$_1$ on Q$_1$ and p$_3$ on Q$_2$ (Fig.~\ref{fig1}\textbf{a} and bottom left panel of Fig.~\ref{fig2}) at a frequency 40 MHz red-detuned from Q$_1$. p$_1$'s amplitude is fixed at $A_\mathrm{XY}/2\pi=65.2~\mathrm{MHz}$, while p$_3$ has parameterized amplitude $A_\mathrm{ZZ}$ and phase $\varphi$. 
For weak driving, $A_\mathrm{XY,ZZ} \ll \omega_\text{q,d}$, the ZZ rate is given as $J_\mathrm{ZZ} \approx 2JA_\mathrm{XY} A_\mathrm{ZZ} \cos(\varphi)/\sqrt{(A_\mathrm{XY}^2+\delta_1^2)(A_\mathrm{ZZ}^2+\delta_2^2)}$,
where $\delta_k$ is the detuning from Q$_k$'s frequency (Methods). Importantly, while the transverse coupling rate $J_\text{XY}$ shown above is essentially limited
by the static coupling strength $J$, the longitudinal coupling rate $J_\mathrm{ZZ}$ can be tuned by two knobs, namely the drives' amplitudes and phase difference.
We characterize the interaction by first initializing the two qubits in the superposition state  $(|0\rangle+|1\rangle)\otimes(|0\rangle+|1\rangle)/2$, applying the pulses as specified, and then extracting the entangling phase $\Phi_\mathrm{ZZ}(\tau_\text{g})=\int_0^{\tau_\text{g}}J_\mathrm{ZZ}\ \!t\ \!dt$ using tomographic reconstruction assisted by numerical optimization  (Methods). As shown in Fig.~\ref{fig2}\textbf{c}, this phase depends on p$_3$'s amplitude $A_\mathrm{ZZ}$ and phase $\varphi$, consistent with our description.

Leveraging the independent controls of the transverse and longitudinal interactions, we now tailor the interplay between them to adjust the anisotropy of the XXZ Heisenberg interaction model. To this end, we apply p$_1$ and p$_2$ pulses with their amplitude $A_\mathrm{XY}$ and duration $\tau_\text{g}$ tuned to induce a full $|10\rangle \leftrightarrow |01\rangle$ swap (Fig.~\ref{fig2}\textbf{b}). Pulse p$_3$ is then jointly applied, albeit with parameterized amplitude $A_\mathrm{ZZ}$ and phase $\varphi$. Incorporating the swap condition into the tomography analysis, we extract the longitudinal entangling phase $\Phi_\mathrm{ZZ}$ which depends on p$_3$'s parameters as shown in Fig.~\ref{fig2}\textbf{d}. This demonstrates that the anisotropy of the model given by Eq.~(\ref{eqn:xxz}) can be programmed in a versatile fashion. 

\vspace{1em}
\noindent\textbf{Benchmarking Heisenberg interactions}


\noindent The programmable Heisenberg interaction endows quantum processors with an extensive quantum gate set as well as the capability to simulate many-body spin-half systems. Here we benchmark our Floquet-engineered Heisenberg interactions by characterizing a suite of representative two-qubit gates: the $\mathrm{iSWAP}$, $\mathrm{CZ}$, and $\mathrm{SWAP}$ gates resulting from the XX Heisenberg model, Ising model, and XXX Heisenberg model, respectively. Accordingly, an $\mathrm{iSWAP}$ unitary arises naturally from a pure transverse $\mathrm{XY}$ interaction with the pulse duration $\tau_\text{g}$ corresponding to a full swap in Fig.~\ref{fig2}\textbf{b}, implemented by applying p$_1$ and p$_2$. In practice, there is a dynamical $\mathrm{ZZ}$ coupling originating from microwave crosstalk, which can be tracked and compensated by simultaneously applying p$_3$ with appropriate amplitude and phase. Likewise, a $\mathrm{CZ}$ gate is realized when p$_1$ and p$_3$ are calibrated to bring up an entangling phase $\Phi_\text{ZZ}/2\pi=0.25$. Finally, we tailor all three pulses to sculpt an isotropic $\mathrm{XXX}$ Heisenberg interaction that leads to a $\mathrm{SWAP}$ gate at the correct gate time $\tau_\text{g}$, at which both $\mathrm{iSWAP}$ and $\mathrm{CZ}$ conditions are satisfied. Our calibration steps are detailed in Methods.

\begin{figure}
    \includegraphics[width=0.5\textwidth]{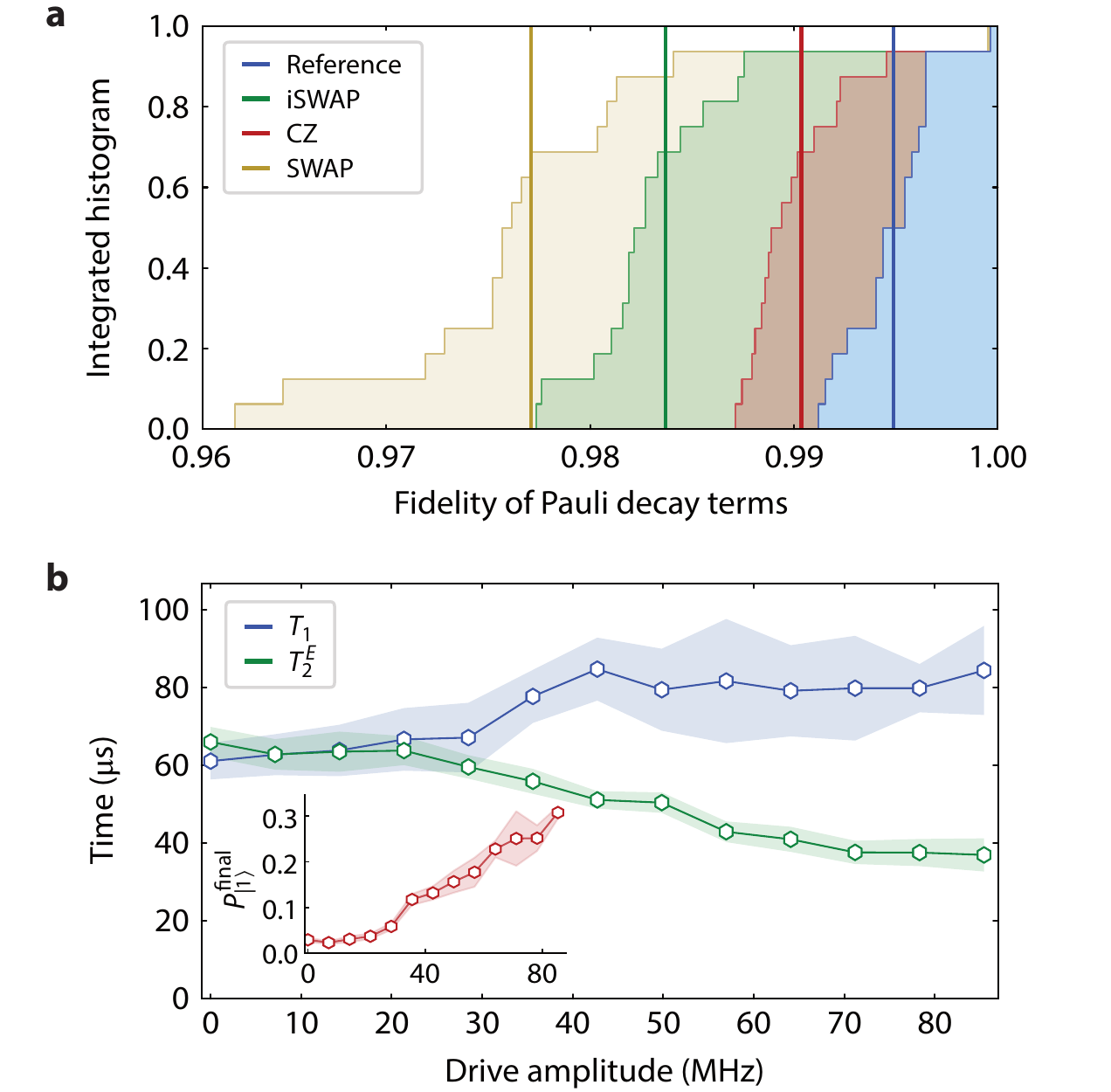}
    \caption{\label{fig3}\textbf{Characterization of two-qubit gates}. \textbf{a.} Cycle benchmarking results showing the Pauli fidelity histogram for the Reference, $\mathrm{iSWAP}$, $\mathrm{CZ}$, and $\mathrm{SWAP}$ cycles, with the solid vertical lines indicating the average values. The corresponding average two-qubit gate fidelities are estimated to be $99.32(3)\%$, $99.72(2)\%$, and $98.93(5)\%$, respectively.  \textbf{b.} Effective coherence time statistics of the Floquet qubit created by adiabatically driving Q$_1$ at a red-detuning of 40 MHz. These are acquired by sampling 20 different sweeps for each curve. Inset: Excited state population at the end of the 355-$\mathrm{\mu s}$-long $|0\rangle \rightarrow |1\rangle$ measurement sequence.}
\end{figure}

To quantify the gates' performances without state preparation and measurement (SPAM) errors, we employ cycle benchmarking (CB)~\cite{erhard2019characterizing}, which tailors all errors into stochastic Pauli channels via Pauli twirling and results in tight bounds on the estimated fidelity  (Methods). In addition to the dressed cycles that include the implemented gates, we also measure the reference cycle and extract its errors to estimate the relevant gate fidelities. Figure~\ref{fig3}\textbf{a} shows the Pauli fidelity distribution histograms of both the reference and dressed cycles corresponding to the intended two-qubit gates. Comparing the dressed cycle data to the reference cycle result allows us to estimate the average gate fidelities of the implemented $\mathrm{iSWAP}$, $\mathrm{CZ}$, and $\mathrm{SWAP}$ gates to be $99.32(3)\%$, $99.72(2)\%$, and $98.93(5)\%$, respectively. We note that these gates are expandable to a continuous fSim gate set~\cite{kivlichan2018quantum,foxen2020demonstrating}, which can be integrated into arbitrary quantum circuits compatible with fixed-frequency qubits by using more advanced circuit compilation tools~\cite{chen2021compiling} or efficient physical Z-gates.

Our analysis attributes the limitations of these results primarily to decoherence mechanisms (Methods). Intriguingly, the Floquet qubits appear to exhibit coherence times deviating from those of the bare qubits, as shown in Fig.~\ref{fig3}\textbf{b}. The measurements are performed using nominal energy relaxation  and echo dephasing procedures on the bare Q$_1$, however, with the addition of a microwave pulse applied 40 MHz red-detuned from its $|0\rangle \leftrightarrow |1\rangle$ transition during idle periods. The results are post-selected to yield the populations of the desired states, and the experiment is repeated over twenty iterations to eliminate any potential outlier. While the dynamics remain the same at small drive amplitudes, $T_1$ tends to increase while $T_2^E$ tends to decrease at strong driving before nonadiabaticity sets in. Interestingly, we also discover a heating mechanism that enlarges the excited state population in the bare qubit at the end of the 355-$\mathrm{\mu s}$-long $|0\rangle\rightarrow |1\rangle$ measurement sequence, with $P_{|1\rangle}^\text{final}$ increases with the driving amplitude (Fig.~\ref{fig3}\textbf{b}, inset). We include additional details in Methods, and hope that future investigations can find efficient approaches to mitigate these effects, reminiscent of the recent progress in driven ultracold atom systems~\cite{viebahn2021suppressing}.


\vspace{1em}
\noindent\textbf{Floquet qutrit and three-qubit gate}

\begin{figure*}
    \includegraphics[width=0.98\textwidth]{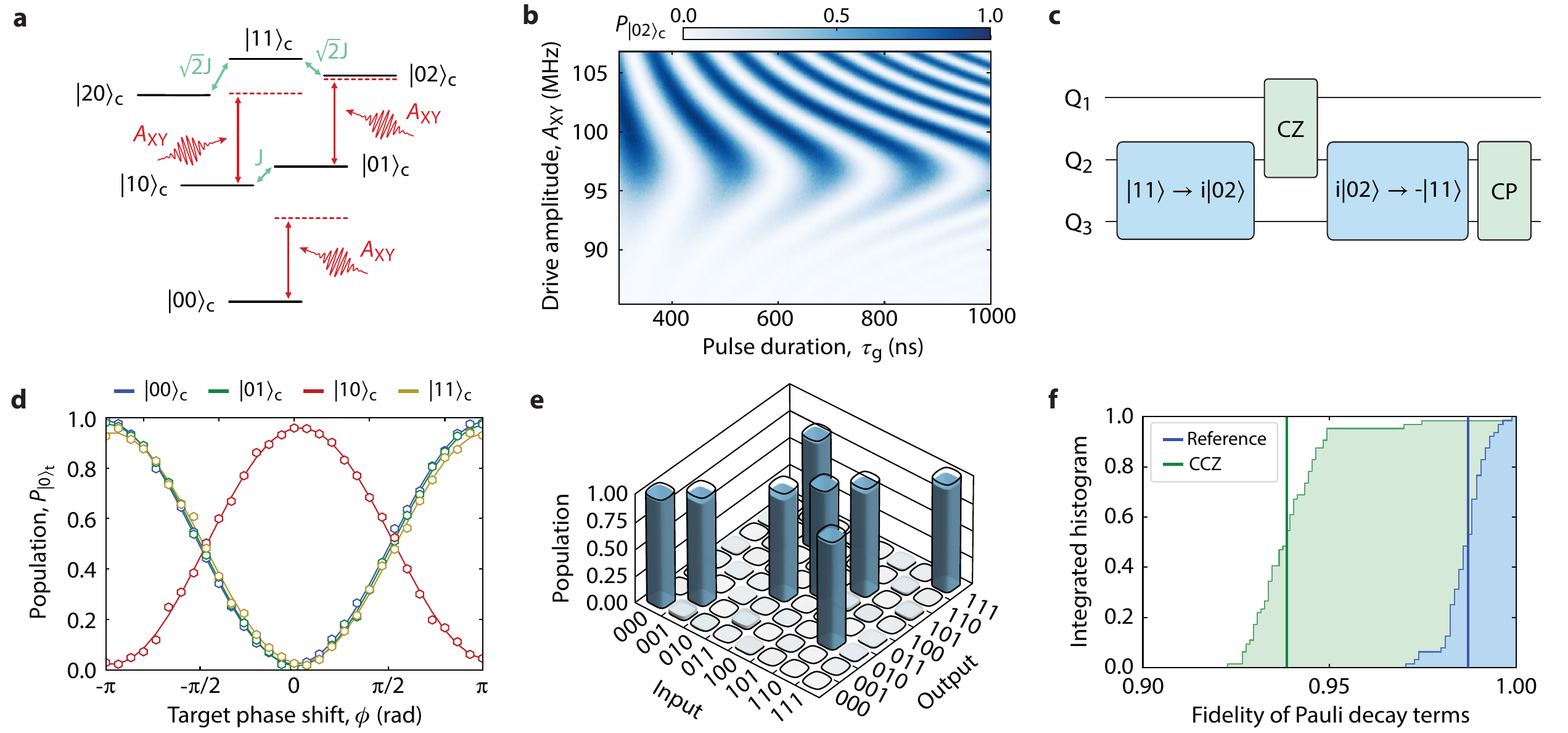}
    \caption{\label{fig4}\textbf{Floquet qutrit and three-qubit CCZ gate}. 
    \textbf{a,} Energy diagram of two coupled transmon circuits Q$_2$ and Q$_3$, with the subscript $\text{c}$ denoting they are control qubits. A microwave drive with amplitude $A_\text{XY}$ is applied at a frequency red-detuned from Q$_3$'s $|1\rangle \leftrightarrow |2\rangle$ transition to create a Floquet qutrit.
    \textbf{b,} Chevron pattern showing a coherent flip-flop between Q$_2$ and Q$_3$'s states $|11\rangle_\text{c}$ and $|02\rangle_\text{c}$, which depends on the amplitude $A_\mathrm{XY}$ and duration $\tau_\text{g}$ of the pulse. The red-detuning is 22 MHz, the ramp time is $\tau_\text{r}=170~\mathrm{ns}$, and the DRAG coefficient is $\lambda_\mathrm{DRAG}=-0.6$.
    \textbf{c,} Gate sequence used to implement a three-qubit $\mathrm{CCZ}$ unitary, with Q$_1$ as the target qubit and Q$_2$, Q$_3$ as the control qubit pair.
    \textbf{d,} A conditionality measurement, using a Ramsey-like sequence \{$\mathrm{R}_\text{Y}^\text{(t)}\!(\frac{\pi}{2})$, $\mathrm{CCZ}$, $\mathrm{R}_\text{Z}^\text{(t)}\!(\phi)$, $\mathrm{R}_\text{Y}^\text{(t)}\!(\frac{\pi}{2})$\}, reveals the dependence of Q$_1$'s phase on the states of Q$_2$ and Q$_3$ under the $\mathrm{CCZ}$ gate implemented using the sequence in \textbf{c}, which is characteristic of a three-body entanglement.
    \textbf{e,} Truth table of the implemented Toffoli gate with a corresponding fidelity of $92.9(1)\%$.
    \textbf{f,} Cycle benchmarking result showing the Pauli fidelity distributions of the three-qubit reference cycle and $\mathrm{CCZ}$ dressed cycle, with the solid vertical lines indicating the average values. The average gate fidelity is estimated to be $96.18(5)\%$.}
\end{figure*}

\noindent So far, the fundamental and universal importance of spin physics motivates our discussion to portray the implemented Floquet qubits as ideal spin-half's. Nevertheless, many solid-state systems, including the transmon, naturally include multiple relevant energy levels. Making use of them expands the Hilbert space, allowing more information to be encoded, which leads to hardware-efficient execution of quantum algorithms~\cite{bocharov2017factoring,gokhale2019asymptotic} and hastens the development of fault-tolerant computation \cite{campbell2014enhanced,muralidharan2017overcoming}. We now show that the presented protocol can be tailored for multi-level systems, thereby paving new pathways for quantum information processing using Floquet qudits. 

Specifically, we leverage the techniques described so far to induce a transverse qutrit-qutrit interaction between the states $|11\rangle$ and $|02\rangle$. Although the cross-Kerr coupling has been explored~\cite{goss2022high}, such an energy-exchange interaction is still absent in fixed-frequency qutrits. While this is a useful ternary gate itself, we presently show that integrating it in a sequence involving multiple qubits allows the implementation of a three-body controlled-controlled-Z gate ($\mathrm{CCZ}$) \cite{fedorov2011implementation}, which plays an important role in quantum applications such as factorization \cite{shor1997polynomial,gidney2021how} and quantum error correction \cite{yoder2016universal,chao2018fault}. To this end, we add to the experiment Q$_3$, which is coupled to the right side of Q$_2$ in Fig.~\ref{fig1}\textbf{a}, and use Q$_2$ and Q$_3$ as control qubits (subscripted $\text{c}$), while Q$_1$ is designated as the target (subscripted $\text{t}$).

Figure~\ref{fig4}\textbf{a} depicts the energy diagram of Q$_2$ and Q$_3$. Our approach to engineer the interaction primarily involves applying a microwave pulse to Q$_3$ at a frequency red-detuned from its $|1\rangle \leftrightarrow |2\rangle$ transition to create a Floquet qutrit such that the control Floquet states $|11\rangle_\text{c}$ and $|02\rangle_\text{c}$ become degenerate. After initializing the control qubits in $|11\rangle_\text{c}$, we apply such a pulse with ramp time $\tau_\text{r}=170~\mathrm{ns}$, DRAG coefficient $\lambda_\mathrm{DRAG}=-0.6$, and varying amplitude $A_\mathrm{XY}$ and duration $\tau_\text{g}$, which are tailored to ensure adiabaticity at a red-detuning of 22 MHz. The transverse interaction between $|11\rangle_\text{c}$ and $|02\rangle_\text{c}$ then manifests into an asymmetric chevron pattern with respect to $A_\mathrm{XY}$ and $\tau_\text{g}$ in Fig.~\ref{fig4}\textbf{b}. Interestingly, we observe that the optimal swap condition occurs at a stronger amplitude relative to the symmetric point. 

A $\mathrm{CCZ}$ unitary can be implemented using the sequence given in Fig.~\ref{fig4}\textbf{c}. The final $\mathrm{CPhase}$ gate on the control qubits are tuned to bring the effective operation on them to be $\hat{I}\otimes\hat{I}$ at the end of the sequence. After calibrating the individual gates, we verify the entanglement between the three qubits by extracting the Z-phase of the target qubit (Q$_1$) for different control states, and observe a phase shift of approximately $\pi$ for $|10\rangle_\text{c}$ (Fig.~\ref{fig4}\textbf{d}), which evinces the $\mathrm{CCZ}$ effect. The sequence can be further sandwiched between single-qubit rotations on Q$_1$ to construct a Toffoli gate. The process can be straightforwardly validated by measuring the truth table, from which we extracted a fidelity of $92.9(1)\%$ (Fig.~\ref{fig4}\textbf{e}). Finally, we employ CB to benchmark the $\mathrm{CCZ}$ gate (Fig.~\ref{fig4}\textbf{f}), achieving a fidelity of $96.18(5)\%$, with the main error resulting from decoherence (Methods).


\vspace{1em}
\noindent\textbf{Outlook}

\noindent Our work embodies a transformative application of Floquet engineering in superconducting circuits where periodic drives are used to map static qubits to Floquet qubits with modifiable quasienergies, granting access to an unconventional tuning channel. We demonstrate the practicality and versatility of this approach by synthesizing Floquet qubits and qutrits, then realizing an XXZ Heisenberg interaction between them with fully tunable anisotropy. The robustness of the scheme is reflected from the high gate fidelities, while the current limitations are straightforward to overcome. On one hand, the coherence times of the fixed-frequency transmon qubits in the experiment are relatively low, therefore we expect better performance in state-of-the-art devices. On the other hand, the coupling rate is primarily limited by the static coupling constant $J$, which can be increased substantially in future devices. In addition, the pulse shape used in this work is quite simple, so we believe that advanced STA techniques can be employed to engineer shorter gates, further reducing errors from dephasing in the future. We note that the full potential of this framework lies upon its adaptation to other synthetic fixed-frequency quantum architectures with better projected performance such as fluxonium quantum processor~\cite{nguyen2022blueprint}.

Having illustrated the useful properties of Floquet qubits and set the stage for immediate improvements, we envision the following avenues to propel the concept in complementary directions. The protocol presented here involves transforming back to the static qubit so normal operations such as readout and single-qubit gates can be employed without recalibration. In the future, a Floquet qubit can be permanently defined by applying a continuous periodic drive in principle, streamlining the process and unlocking new opportunities for novel control and readout methods~\cite{gandon2022engineering}. This approach also allows \textit{in-situ} tuning of the qubit frequencies, thus provides a practical solution for problems arising from two-level-system defects and spectral crowding. In another route, the ramp time can be reduced substantially if we operate in the diabatic regime, where the mapping is close to ideal despite finite transition between the Floquet states. We expect the potential development of optimal control within the Floquet framework to provide a reliable approach in this regime. Last but not least, the heating effect which correlates with the reduction in $T_2$ reminisces a similar effect in cold atom systems which has been successfully suppressed~\cite{viebahn2021suppressing}. This calls for deeper understanding of the quantum thermodynamics in driven solid-state systems and possible mitigation strategies in the future.

\section*{Acknowledgments}
We thank Alexis Morvan and William P. Livingston for measurement assistance. L.B.N is grateful to Ian Mondragon-Shem and Jens Koch for valuable discussions on Floquet theory. The micrograph of the chip was obtained with the support from Bingcheng Qing and Kyunghoon Lee. This work was supported by the Office of Advanced Scientific Computing Research, Testbeds for Science program, Office of Science of the U.S. Department of Energy under Contract No. DE-AC02-05CH11231, the KIST research program under grant No. 2E31531, and ARO grant No. W911NF-22-1-0258.



    
\newpage
\bibliography{apssamp}

\begin{thebibliography}{52}%
\makeatletter
\providecommand \@ifxundefined [1]{%
 \@ifx{#1\undefined}
}%
\providecommand \@ifnum [1]{%
 \ifnum #1\expandafter \@firstoftwo
 \else \expandafter \@secondoftwo
 \fi
}%
\providecommand \@ifx [1]{%
 \ifx #1\expandafter \@firstoftwo
 \else \expandafter \@secondoftwo
 \fi
}%
\providecommand \natexlab [1]{#1}%
\providecommand \enquote  [1]{``#1''}%
\providecommand \bibnamefont  [1]{#1}%
\providecommand \bibfnamefont [1]{#1}%
\providecommand \citenamefont [1]{#1}%
\providecommand \href@noop [0]{\@secondoftwo}%
\providecommand \href [0]{\begingroup \@sanitize@url \@href}%
\providecommand \@href[1]{\@@startlink{#1}\@@href}%
\providecommand \@@href[1]{\endgroup#1\@@endlink}%
\providecommand \@sanitize@url [0]{\catcode `\\12\catcode `\$12\catcode
  `\&12\catcode `\#12\catcode `\^12\catcode `\_12\catcode `\%12\relax}%
\providecommand \@@startlink[1]{}%
\providecommand \@@endlink[0]{}%
\providecommand \url  [0]{\begingroup\@sanitize@url \@url }%
\providecommand \@url [1]{\endgroup\@href {#1}{\urlprefix }}%
\providecommand \urlprefix  [0]{URL }%
\providecommand \Eprint [0]{\href }%
\providecommand \doibase [0]{http://dx.doi.org/}%
\providecommand \selectlanguage [0]{\@gobble}%
\providecommand \bibinfo  [0]{\@secondoftwo}%
\providecommand \bibfield  [0]{\@secondoftwo}%
\providecommand \translation [1]{[#1]}%
\providecommand \BibitemOpen [0]{}%
\providecommand \bibitemStop [0]{}%
\providecommand \bibitemNoStop [0]{.\EOS\space}%
\providecommand \EOS [0]{\spacefactor3000\relax}%
\providecommand \BibitemShut  [1]{\csname bibitem#1\endcsname}%
\let\auto@bib@innerbib\@empty
\bibitem [{\citenamefont {Gyenis}\ \emph {et~al.}(2021)\citenamefont {Gyenis},
  \citenamefont {Di~Paolo}, \citenamefont {Koch}, \citenamefont {Blais},
  \citenamefont {Houck},\ and\ \citenamefont {Schuster}}]{gyenis2021moving}%
  \BibitemOpen
  \bibfield  {author} {\bibinfo {author} {\bibfnamefont {A.}~\bibnamefont
  {Gyenis}}, \bibinfo {author} {\bibfnamefont {A.}~\bibnamefont {Di~Paolo}},
  \bibinfo {author} {\bibfnamefont {J.}~\bibnamefont {Koch}}, \bibinfo {author}
  {\bibfnamefont {A.}~\bibnamefont {Blais}}, \bibinfo {author} {\bibfnamefont
  {A.~A.}\ \bibnamefont {Houck}}, \ and\ \bibinfo {author} {\bibfnamefont
  {D.~I.}\ \bibnamefont {Schuster}},\ }\href {\doibase
  10.1103/PRXQuantum.2.030101} {\bibfield  {journal} {\bibinfo  {journal} {PRX
  Quantum}\ }\textbf {\bibinfo {volume} {2}},\ \bibinfo {pages} {030101}
  (\bibinfo {year} {2021})}\BibitemShut {NoStop}%
\bibitem [{\citenamefont {Siddiqi}(2021)}]{siddiqi2021engineering}%
  \BibitemOpen
  \bibfield  {author} {\bibinfo {author} {\bibfnamefont {I.}~\bibnamefont
  {Siddiqi}},\ }\href {\doibase 10.1038/s41578-021-00370-4} {\bibfield
  {journal} {\bibinfo  {journal} {Nat. Rev. Mater.}\ }\textbf {\bibinfo
  {volume} {6}},\ \bibinfo {pages} {875} (\bibinfo {year} {2021})}\BibitemShut
  {NoStop}%
\bibitem [{\citenamefont {Savary}\ and\ \citenamefont
  {Balents}(2016)}]{savary2016quantum}%
  \BibitemOpen
  \bibfield  {author} {\bibinfo {author} {\bibfnamefont {L.}~\bibnamefont
  {Savary}}\ and\ \bibinfo {author} {\bibfnamefont {L.}~\bibnamefont
  {Balents}},\ }\href {\doibase 10.1088/0034-4885/80/1/016502} {\bibfield
  {journal} {\bibinfo  {journal} {Rep. Prog. Phys.}\ }\textbf {\bibinfo
  {volume} {80}},\ \bibinfo {pages} {016502} (\bibinfo {year}
  {2016})}\BibitemShut {NoStop}%
\bibitem [{\citenamefont {Kivlichan}\ \emph {et~al.}(2018)\citenamefont
  {Kivlichan}, \citenamefont {McClean}, \citenamefont {Wiebe}, \citenamefont
  {Gidney}, \citenamefont {Aspuru-Guzik}, \citenamefont {Chan},\ and\
  \citenamefont {Babbush}}]{kivlichan2018quantum}%
  \BibitemOpen
  \bibfield  {author} {\bibinfo {author} {\bibfnamefont {I.~D.}\ \bibnamefont
  {Kivlichan}}, \bibinfo {author} {\bibfnamefont {J.}~\bibnamefont {McClean}},
  \bibinfo {author} {\bibfnamefont {N.}~\bibnamefont {Wiebe}}, \bibinfo
  {author} {\bibfnamefont {C.}~\bibnamefont {Gidney}}, \bibinfo {author}
  {\bibfnamefont {A.}~\bibnamefont {Aspuru-Guzik}}, \bibinfo {author}
  {\bibfnamefont {G.~K.-L.}\ \bibnamefont {Chan}}, \ and\ \bibinfo {author}
  {\bibfnamefont {R.}~\bibnamefont {Babbush}},\ }\href {\doibase
  10.1103/PhysRevLett.120.110501} {\bibfield  {journal} {\bibinfo  {journal}
  {Phys. Rev. Lett.}\ }\textbf {\bibinfo {volume} {120}},\ \bibinfo {pages}
  {110501} (\bibinfo {year} {2018})}\BibitemShut {NoStop}%
\bibitem [{\citenamefont {Shirley}(1965)}]{shirley1965solution}%
  \BibitemOpen
  \bibfield  {author} {\bibinfo {author} {\bibfnamefont {J.~H.}\ \bibnamefont
  {Shirley}},\ }\href {\doibase 10.1103/PhysRev.138.B979} {\bibfield  {journal}
  {\bibinfo  {journal} {Phys. Rev.}\ }\textbf {\bibinfo {volume} {138}},\
  \bibinfo {pages} {B979} (\bibinfo {year} {1965})}\BibitemShut {NoStop}%
\bibitem [{\citenamefont {Bernien}\ \emph {et~al.}(2017)\citenamefont {Bernien}
  \emph {et~al.}}]{bernien2017probing}%
  \BibitemOpen
  \bibfield  {author} {\bibinfo {author} {\bibfnamefont {H.}~\bibnamefont
  {Bernien}} \emph {et~al.},\ }\href {\doibase 10.1038/nature24622} {\bibfield
  {journal} {\bibinfo  {journal} {Nature}\ }\textbf {\bibinfo {volume} {551}},\
  \bibinfo {pages} {579} (\bibinfo {year} {2017})}\BibitemShut {NoStop}%
\bibitem [{\citenamefont {Jepsen}\ \emph {et~al.}(2020)\citenamefont {Jepsen},
  \citenamefont {Amato-Grill}, \citenamefont {Dimitrova}, \citenamefont {Ho},
  \citenamefont {Demler},\ and\ \citenamefont {Ketterle}}]{jepsen2020spin}%
  \BibitemOpen
  \bibfield  {author} {\bibinfo {author} {\bibfnamefont {P.~N.}\ \bibnamefont
  {Jepsen}}, \bibinfo {author} {\bibfnamefont {J.}~\bibnamefont {Amato-Grill}},
  \bibinfo {author} {\bibfnamefont {I.}~\bibnamefont {Dimitrova}}, \bibinfo
  {author} {\bibfnamefont {W.~W.}\ \bibnamefont {Ho}}, \bibinfo {author}
  {\bibfnamefont {E.}~\bibnamefont {Demler}}, \ and\ \bibinfo {author}
  {\bibfnamefont {W.}~\bibnamefont {Ketterle}},\ }\href {\doibase
  10.1038/s41586-020-3033-y} {\bibfield  {journal} {\bibinfo  {journal}
  {Nature}\ }\textbf {\bibinfo {volume} {588}},\ \bibinfo {pages} {403}
  (\bibinfo {year} {2020})}\BibitemShut {NoStop}%
\bibitem [{\citenamefont {Zhang}\ \emph {et~al.}(2017)\citenamefont {Zhang}
  \emph {et~al.}}]{zhang2017observation}%
  \BibitemOpen
  \bibfield  {author} {\bibinfo {author} {\bibfnamefont {J.}~\bibnamefont
  {Zhang}} \emph {et~al.},\ }\href {\doibase 10.1038/nature21413} {\bibfield
  {journal} {\bibinfo  {journal} {Nature}\ }\textbf {\bibinfo {volume} {543}},\
  \bibinfo {pages} {217} (\bibinfo {year} {2017})}\BibitemShut {NoStop}%
\bibitem [{\citenamefont {Jepsen}\ \emph {et~al.}(2022)\citenamefont {Jepsen},
  \citenamefont {Lee}, \citenamefont {Lin}, \citenamefont {Dimitrova},
  \citenamefont {Margalit}, \citenamefont {Ho},\ and\ \citenamefont
  {Ketterle}}]{jepsen2022long}%
  \BibitemOpen
  \bibfield  {author} {\bibinfo {author} {\bibfnamefont {P.~N.}\ \bibnamefont
  {Jepsen}}, \bibinfo {author} {\bibfnamefont {Y.~K.}\ \bibnamefont {Lee}},
  \bibinfo {author} {\bibfnamefont {H.}~\bibnamefont {Lin}}, \bibinfo {author}
  {\bibfnamefont {I.}~\bibnamefont {Dimitrova}}, \bibinfo {author}
  {\bibfnamefont {Y.}~\bibnamefont {Margalit}}, \bibinfo {author}
  {\bibfnamefont {W.~W.}\ \bibnamefont {Ho}}, \ and\ \bibinfo {author}
  {\bibfnamefont {W.}~\bibnamefont {Ketterle}},\ }\href {\doibase
  10.1038/s41567-022-01651-7} {\bibfield  {journal} {\bibinfo  {journal} {Nat.
  Phys.}\ }\textbf {\bibinfo {volume} {18}},\ \bibinfo {pages} {899} (\bibinfo
  {year} {2022})}\BibitemShut {NoStop}%
\bibitem [{\citenamefont {Morvan}\ \emph {et~al.}(2022)\citenamefont {Morvan}
  \emph {et~al.}}]{morvan2022formation}%
  \BibitemOpen
  \bibfield  {author} {\bibinfo {author} {\bibfnamefont {A.}~\bibnamefont
  {Morvan}} \emph {et~al.},\ }\href@noop {} {\bibfield  {journal} {\bibinfo
  {journal} {Preprint at \url{https://arxiv.org/abs/2206.05254}}\ } (\bibinfo
  {year} {2022})}\BibitemShut {NoStop}%
\bibitem [{\citenamefont {Bharti}\ \emph {et~al.}(2022)\citenamefont {Bharti}
  \emph {et~al.}}]{bharti2022noisy}%
  \BibitemOpen
  \bibfield  {author} {\bibinfo {author} {\bibfnamefont {K.}~\bibnamefont
  {Bharti}} \emph {et~al.},\ }\href {\doibase 10.1103/RevModPhys.94.015004}
  {\bibfield  {journal} {\bibinfo  {journal} {Rev. Mod. Phys.}\ }\textbf
  {\bibinfo {volume} {94}},\ \bibinfo {pages} {015004} (\bibinfo {year}
  {2022})}\BibitemShut {NoStop}%
\bibitem [{\citenamefont {Fowler}\ \emph {et~al.}(2012)\citenamefont {Fowler},
  \citenamefont {Mariantoni}, \citenamefont {Martinis},\ and\ \citenamefont
  {Cleland}}]{fowler2012surface}%
  \BibitemOpen
  \bibfield  {author} {\bibinfo {author} {\bibfnamefont {A.~G.}\ \bibnamefont
  {Fowler}}, \bibinfo {author} {\bibfnamefont {M.}~\bibnamefont {Mariantoni}},
  \bibinfo {author} {\bibfnamefont {J.~M.}\ \bibnamefont {Martinis}}, \ and\
  \bibinfo {author} {\bibfnamefont {A.~N.}\ \bibnamefont {Cleland}},\ }\href
  {\doibase 10.1103/PhysRevA.86.032324} {\bibfield  {journal} {\bibinfo
  {journal} {Phys. Rev. A}\ }\textbf {\bibinfo {volume} {86}},\ \bibinfo
  {pages} {032324} (\bibinfo {year} {2012})}\BibitemShut {NoStop}%
\bibitem [{\citenamefont {Ghosh}\ and\ \citenamefont
  {Fowler}(2015)}]{ghosh2015leakage}%
  \BibitemOpen
  \bibfield  {author} {\bibinfo {author} {\bibfnamefont {J.}~\bibnamefont
  {Ghosh}}\ and\ \bibinfo {author} {\bibfnamefont {A.~G.}\ \bibnamefont
  {Fowler}},\ }\href {\doibase 10.1103/PhysRevA.91.020302} {\bibfield
  {journal} {\bibinfo  {journal} {Phys. Rev. A}\ }\textbf {\bibinfo {volume}
  {91}},\ \bibinfo {pages} {020302} (\bibinfo {year} {2015})}\BibitemShut
  {NoStop}%
\bibitem [{\citenamefont {Koch}\ \emph {et~al.}(2007)\citenamefont {Koch} \emph
  {et~al.}}]{koch2007charge}%
  \BibitemOpen
  \bibfield  {author} {\bibinfo {author} {\bibfnamefont {J.}~\bibnamefont
  {Koch}} \emph {et~al.},\ }\href {\doibase 10.1103/PhysRevA.76.042319}
  {\bibfield  {journal} {\bibinfo  {journal} {Phys. Rev. A}\ }\textbf {\bibinfo
  {volume} {76}},\ \bibinfo {pages} {042319} (\bibinfo {year}
  {2007})}\BibitemShut {NoStop}%
\bibitem [{\citenamefont {Paik}\ \emph {et~al.}(2011)\citenamefont {Paik} \emph
  {et~al.}}]{paik2011observation}%
  \BibitemOpen
  \bibfield  {author} {\bibinfo {author} {\bibfnamefont {H.}~\bibnamefont
  {Paik}} \emph {et~al.},\ }\href {\doibase 10.1103/PhysRevLett.107.240501}
  {\bibfield  {journal} {\bibinfo  {journal} {Phys. Rev. Lett.}\ }\textbf
  {\bibinfo {volume} {107}},\ \bibinfo {pages} {240501} (\bibinfo {year}
  {2011})}\BibitemShut {NoStop}%
\bibitem [{\citenamefont {Manucharyan}\ \emph {et~al.}(2009)\citenamefont
  {Manucharyan}, \citenamefont {Koch}, \citenamefont {Glazman},\ and\
  \citenamefont {Devoret}}]{manucharyan2009fluxonium}%
  \BibitemOpen
  \bibfield  {author} {\bibinfo {author} {\bibfnamefont {V.~E.}\ \bibnamefont
  {Manucharyan}}, \bibinfo {author} {\bibfnamefont {J.}~\bibnamefont {Koch}},
  \bibinfo {author} {\bibfnamefont {L.~I.}\ \bibnamefont {Glazman}}, \ and\
  \bibinfo {author} {\bibfnamefont {M.~H.}\ \bibnamefont {Devoret}},\ }\href
  {\doibase 10.1126/science.1175552} {\bibfield  {journal} {\bibinfo  {journal}
  {Science}\ }\textbf {\bibinfo {volume} {326}},\ \bibinfo {pages} {113}
  (\bibinfo {year} {2009})}\BibitemShut {NoStop}%
\bibitem [{\citenamefont {Nguyen}\ \emph {et~al.}(2019)\citenamefont {Nguyen},
  \citenamefont {Lin}, \citenamefont {Somoroff}, \citenamefont {Mencia},
  \citenamefont {Grabon},\ and\ \citenamefont {Manucharyan}}]{nguyen2019high}%
  \BibitemOpen
  \bibfield  {author} {\bibinfo {author} {\bibfnamefont {L.~B.}\ \bibnamefont
  {Nguyen}}, \bibinfo {author} {\bibfnamefont {Y.-H.}\ \bibnamefont {Lin}},
  \bibinfo {author} {\bibfnamefont {A.}~\bibnamefont {Somoroff}}, \bibinfo
  {author} {\bibfnamefont {R.}~\bibnamefont {Mencia}}, \bibinfo {author}
  {\bibfnamefont {N.}~\bibnamefont {Grabon}}, \ and\ \bibinfo {author}
  {\bibfnamefont {V.~E.}\ \bibnamefont {Manucharyan}},\ }\href {\doibase
  10.1103/PhysRevX.9.041041} {\bibfield  {journal} {\bibinfo  {journal} {Phys.
  Rev. X}\ }\textbf {\bibinfo {volume} {9}},\ \bibinfo {pages} {041041}
  (\bibinfo {year} {2019})}\BibitemShut {NoStop}%
\bibitem [{\citenamefont {Chow}\ \emph {et~al.}(2011)\citenamefont {Chow} \emph
  {et~al.}}]{chow2011simple}%
  \BibitemOpen
  \bibfield  {author} {\bibinfo {author} {\bibfnamefont {J.~M.}\ \bibnamefont
  {Chow}} \emph {et~al.},\ }\href {\doibase 10.1103/PhysRevLett.107.080502}
  {\bibfield  {journal} {\bibinfo  {journal} {Phys. Rev. Lett.}\ }\textbf
  {\bibinfo {volume} {107}},\ \bibinfo {pages} {080502} (\bibinfo {year}
  {2011})}\BibitemShut {NoStop}%
\bibitem [{\citenamefont {Kim}\ \emph {et~al.}(2022)\citenamefont {Kim} \emph
  {et~al.}}]{kim2022high}%
  \BibitemOpen
  \bibfield  {author} {\bibinfo {author} {\bibfnamefont {Y.}~\bibnamefont
  {Kim}} \emph {et~al.},\ }\href {\doibase 10.1038/s41567-022-01590-3}
  {\bibfield  {journal} {\bibinfo  {journal} {Nat. Phys.}\ }\textbf {\bibinfo
  {volume} {18}},\ \bibinfo {pages} {783} (\bibinfo {year} {2022})}\BibitemShut
  {NoStop}%
\bibitem [{\citenamefont {Mitchell}\ \emph {et~al.}(2021)\citenamefont
  {Mitchell} \emph {et~al.}}]{mitchell2021hardware}%
  \BibitemOpen
  \bibfield  {author} {\bibinfo {author} {\bibfnamefont {B.~K.}\ \bibnamefont
  {Mitchell}} \emph {et~al.},\ }\href {\doibase 10.1103/PhysRevLett.127.200502}
  {\bibfield  {journal} {\bibinfo  {journal} {Phys. Rev. Lett.}\ }\textbf
  {\bibinfo {volume} {127}},\ \bibinfo {pages} {200502} (\bibinfo {year}
  {2021})}\BibitemShut {NoStop}%
\bibitem [{\citenamefont {Wei}\ \emph {et~al.}(2022)\citenamefont {Wei} \emph
  {et~al.}}]{wei2022hamiltonian}%
  \BibitemOpen
  \bibfield  {author} {\bibinfo {author} {\bibfnamefont {K.~X.}\ \bibnamefont
  {Wei}} \emph {et~al.},\ }\href {\doibase 10.1103/PhysRevLett.129.060501}
  {\bibfield  {journal} {\bibinfo  {journal} {Phys. Rev. Lett.}\ }\textbf
  {\bibinfo {volume} {129}},\ \bibinfo {pages} {060501} (\bibinfo {year}
  {2022})}\BibitemShut {NoStop}%
\bibitem [{\citenamefont {Ganzhorn}\ \emph {et~al.}(2020)\citenamefont
  {Ganzhorn} \emph {et~al.}}]{ganzhorn2020benchmarking}%
  \BibitemOpen
  \bibfield  {author} {\bibinfo {author} {\bibfnamefont {M.}~\bibnamefont
  {Ganzhorn}} \emph {et~al.},\ }\href {\doibase
  10.1103/PhysRevResearch.2.033447} {\bibfield  {journal} {\bibinfo  {journal}
  {Phys. Rev. Research}\ }\textbf {\bibinfo {volume} {2}},\ \bibinfo {pages}
  {033447} (\bibinfo {year} {2020})}\BibitemShut {NoStop}%
\bibitem [{\citenamefont {Huang}\ \emph {et~al.}(2021)\citenamefont {Huang},
  \citenamefont {Mundada}, \citenamefont {Gyenis}, \citenamefont {Schuster},
  \citenamefont {Houck},\ and\ \citenamefont {Koch}}]{huang2021engineering}%
  \BibitemOpen
  \bibfield  {author} {\bibinfo {author} {\bibfnamefont {Z.}~\bibnamefont
  {Huang}}, \bibinfo {author} {\bibfnamefont {P.~S.}\ \bibnamefont {Mundada}},
  \bibinfo {author} {\bibfnamefont {A.}~\bibnamefont {Gyenis}}, \bibinfo
  {author} {\bibfnamefont {D.~I.}\ \bibnamefont {Schuster}}, \bibinfo {author}
  {\bibfnamefont {A.~A.}\ \bibnamefont {Houck}}, \ and\ \bibinfo {author}
  {\bibfnamefont {J.}~\bibnamefont {Koch}},\ }\href {\doibase
  10.1103/PhysRevApplied.15.034065} {\bibfield  {journal} {\bibinfo  {journal}
  {Phys. Rev. Applied}\ }\textbf {\bibinfo {volume} {15}},\ \bibinfo {pages}
  {034065} (\bibinfo {year} {2021})}\BibitemShut {NoStop}%
\bibitem [{\citenamefont {Gandon}\ \emph {et~al.}(2022)\citenamefont {Gandon},
  \citenamefont {Le~Calonnec}, \citenamefont {Shillito}, \citenamefont
  {Petrescu},\ and\ \citenamefont {Blais}}]{gandon2022engineering}%
  \BibitemOpen
  \bibfield  {author} {\bibinfo {author} {\bibfnamefont {A.}~\bibnamefont
  {Gandon}}, \bibinfo {author} {\bibfnamefont {C.}~\bibnamefont {Le~Calonnec}},
  \bibinfo {author} {\bibfnamefont {R.}~\bibnamefont {Shillito}}, \bibinfo
  {author} {\bibfnamefont {A.}~\bibnamefont {Petrescu}}, \ and\ \bibinfo
  {author} {\bibfnamefont {A.}~\bibnamefont {Blais}},\ }\href {\doibase
  10.1103/PhysRevApplied.17.064006} {\bibfield  {journal} {\bibinfo  {journal}
  {Phys. Rev. Applied}\ }\textbf {\bibinfo {volume} {17}},\ \bibinfo {pages}
  {064006} (\bibinfo {year} {2022})}\BibitemShut {NoStop}%
\bibitem [{\citenamefont {Gu\'ery-Odelin}\ \emph {et~al.}(2019)\citenamefont
  {Gu\'ery-Odelin}, \citenamefont {Ruschhaupt}, \citenamefont {Kiely},
  \citenamefont {Torrontegui}, \citenamefont {Mart\'{\i}nez-Garaot},\ and\
  \citenamefont {Muga}}]{gu2019shortcuts}%
  \BibitemOpen
  \bibfield  {author} {\bibinfo {author} {\bibfnamefont {D.}~\bibnamefont
  {Gu\'ery-Odelin}}, \bibinfo {author} {\bibfnamefont {A.}~\bibnamefont
  {Ruschhaupt}}, \bibinfo {author} {\bibfnamefont {A.}~\bibnamefont {Kiely}},
  \bibinfo {author} {\bibfnamefont {E.}~\bibnamefont {Torrontegui}}, \bibinfo
  {author} {\bibfnamefont {S.}~\bibnamefont {Mart\'{\i}nez-Garaot}}, \ and\
  \bibinfo {author} {\bibfnamefont {J.~G.}\ \bibnamefont {Muga}},\ }\href
  {\doibase 10.1103/RevModPhys.91.045001} {\bibfield  {journal} {\bibinfo
  {journal} {Rev. Mod. Phys.}\ }\textbf {\bibinfo {volume} {91}},\ \bibinfo
  {pages} {045001} (\bibinfo {year} {2019})}\BibitemShut {NoStop}%
\bibitem [{\citenamefont {Motzoi}\ \emph {et~al.}(2009)\citenamefont {Motzoi},
  \citenamefont {Gambetta}, \citenamefont {Rebentrost},\ and\ \citenamefont
  {Wilhelm}}]{motzoi2009simple}%
  \BibitemOpen
  \bibfield  {author} {\bibinfo {author} {\bibfnamefont {F.}~\bibnamefont
  {Motzoi}}, \bibinfo {author} {\bibfnamefont {J.~M.}\ \bibnamefont
  {Gambetta}}, \bibinfo {author} {\bibfnamefont {P.}~\bibnamefont
  {Rebentrost}}, \ and\ \bibinfo {author} {\bibfnamefont {F.~K.}\ \bibnamefont
  {Wilhelm}},\ }\href {\doibase 10.1103/PhysRevLett.103.110501} {\bibfield
  {journal} {\bibinfo  {journal} {Phys. Rev. Lett.}\ }\textbf {\bibinfo
  {volume} {103}},\ \bibinfo {pages} {110501} (\bibinfo {year}
  {2009})}\BibitemShut {NoStop}%
\bibitem [{\citenamefont {Gambetta}\ \emph {et~al.}(2011)\citenamefont
  {Gambetta}, \citenamefont {Motzoi}, \citenamefont {Merkel},\ and\
  \citenamefont {Wilhelm}}]{gambetta2011analytic}%
  \BibitemOpen
  \bibfield  {author} {\bibinfo {author} {\bibfnamefont {J.~M.}\ \bibnamefont
  {Gambetta}}, \bibinfo {author} {\bibfnamefont {F.}~\bibnamefont {Motzoi}},
  \bibinfo {author} {\bibfnamefont {S.~T.}\ \bibnamefont {Merkel}}, \ and\
  \bibinfo {author} {\bibfnamefont {F.~K.}\ \bibnamefont {Wilhelm}},\ }\href
  {\doibase 10.1103/PhysRevA.83.012308} {\bibfield  {journal} {\bibinfo
  {journal} {Phys. Rev. A}\ }\textbf {\bibinfo {volume} {83}},\ \bibinfo
  {pages} {012308} (\bibinfo {year} {2011})}\BibitemShut {NoStop}%
\bibitem [{\citenamefont {Fedorov}\ \emph {et~al.}(2011)\citenamefont
  {Fedorov}, \citenamefont {Steffen}, \citenamefont {Baur}, \citenamefont
  {da~Silva},\ and\ \citenamefont {Wallraff}}]{fedorov2011implementation}%
  \BibitemOpen
  \bibfield  {author} {\bibinfo {author} {\bibfnamefont {A.}~\bibnamefont
  {Fedorov}}, \bibinfo {author} {\bibfnamefont {L.}~\bibnamefont {Steffen}},
  \bibinfo {author} {\bibfnamefont {M.}~\bibnamefont {Baur}}, \bibinfo {author}
  {\bibfnamefont {M.~P.}\ \bibnamefont {da~Silva}}, \ and\ \bibinfo {author}
  {\bibfnamefont {A.}~\bibnamefont {Wallraff}},\ }\href {\doibase
  10.1038/nature10713} {\bibfield  {journal} {\bibinfo  {journal} {Nature}\
  }\textbf {\bibinfo {volume} {481}},\ \bibinfo {pages} {170} (\bibinfo {year}
  {2011})}\BibitemShut {NoStop}%
\bibitem [{\citenamefont {Majer}\ \emph {et~al.}(2007)\citenamefont {Majer}
  \emph {et~al.}}]{majer2007coupling}%
  \BibitemOpen
  \bibfield  {author} {\bibinfo {author} {\bibfnamefont {J.}~\bibnamefont
  {Majer}} \emph {et~al.},\ }\href {\doibase 10.1038/nature06184} {\bibfield
  {journal} {\bibinfo  {journal} {Nature}\ }\textbf {\bibinfo {volume} {449}},\
  \bibinfo {pages} {443} (\bibinfo {year} {2007})}\BibitemShut {NoStop}%
\bibitem [{\citenamefont {Blais}\ \emph {et~al.}(2021)\citenamefont {Blais},
  \citenamefont {Grimsmo}, \citenamefont {Girvin},\ and\ \citenamefont
  {Wallraff}}]{blais2021circuit}%
  \BibitemOpen
  \bibfield  {author} {\bibinfo {author} {\bibfnamefont {A.}~\bibnamefont
  {Blais}}, \bibinfo {author} {\bibfnamefont {A.~L.}\ \bibnamefont {Grimsmo}},
  \bibinfo {author} {\bibfnamefont {S.~M.}\ \bibnamefont {Girvin}}, \ and\
  \bibinfo {author} {\bibfnamefont {A.}~\bibnamefont {Wallraff}},\ }\href
  {\doibase 10.1103/RevModPhys.93.025005} {\bibfield  {journal} {\bibinfo
  {journal} {Rev. Mod. Phys.}\ }\textbf {\bibinfo {volume} {93}},\ \bibinfo
  {pages} {025005} (\bibinfo {year} {2021})}\BibitemShut {NoStop}%
\bibitem [{\citenamefont {Weitenberg}\ and\ \citenamefont
  {Simonet}(2021)}]{weitenberg2021tailoring}%
  \BibitemOpen
  \bibfield  {author} {\bibinfo {author} {\bibfnamefont {C.}~\bibnamefont
  {Weitenberg}}\ and\ \bibinfo {author} {\bibfnamefont {J.}~\bibnamefont
  {Simonet}},\ }\href {\doibase 10.1038/s41567-021-01316-x} {\bibfield
  {journal} {\bibinfo  {journal} {Nat. Phys.}\ }\textbf {\bibinfo {volume}
  {17}},\ \bibinfo {pages} {1342} (\bibinfo {year} {2021})}\BibitemShut
  {NoStop}%
\bibitem [{\citenamefont {Deng}\ \emph {et~al.}(2016)\citenamefont {Deng},
  \citenamefont {Shen}, \citenamefont {Ashhab},\ and\ \citenamefont
  {Lupascu}}]{deng2016dynamics}%
  \BibitemOpen
  \bibfield  {author} {\bibinfo {author} {\bibfnamefont {C.}~\bibnamefont
  {Deng}}, \bibinfo {author} {\bibfnamefont {F.}~\bibnamefont {Shen}}, \bibinfo
  {author} {\bibfnamefont {S.}~\bibnamefont {Ashhab}}, \ and\ \bibinfo {author}
  {\bibfnamefont {A.}~\bibnamefont {Lupascu}},\ }\href {\doibase
  10.1103/PhysRevA.94.032323} {\bibfield  {journal} {\bibinfo  {journal} {Phys.
  Rev. A}\ }\textbf {\bibinfo {volume} {94}},\ \bibinfo {pages} {032323}
  (\bibinfo {year} {2016})}\BibitemShut {NoStop}%
\bibitem [{\citenamefont {Erhard}\ \emph {et~al.}(2019)\citenamefont {Erhard}
  \emph {et~al.}}]{erhard2019characterizing}%
  \BibitemOpen
  \bibfield  {author} {\bibinfo {author} {\bibfnamefont {A.}~\bibnamefont
  {Erhard}} \emph {et~al.},\ }\href {\doibase 10.1038/s41467-019-13068-7}
  {\bibfield  {journal} {\bibinfo  {journal} {Nat. Commun.}\ }\textbf {\bibinfo
  {volume} {10}},\ \bibinfo {pages} {1} (\bibinfo {year} {2019})}\BibitemShut
  {NoStop}%
\bibitem [{\citenamefont {Foxen}\ \emph {et~al.}(2020)\citenamefont {Foxen}
  \emph {et~al.}}]{foxen2020demonstrating}%
  \BibitemOpen
  \bibfield  {author} {\bibinfo {author} {\bibfnamefont {B.}~\bibnamefont
  {Foxen}} \emph {et~al.},\ }\href {\doibase 10.1103/PhysRevLett.125.120504}
  {\bibfield  {journal} {\bibinfo  {journal} {Phys. Rev. Lett.}\ }\textbf
  {\bibinfo {volume} {125}},\ \bibinfo {pages} {120504} (\bibinfo {year}
  {2020})}\BibitemShut {NoStop}%
\bibitem [{\citenamefont {Chen}\ \emph {et~al.}(2021)\citenamefont {Chen},
  \citenamefont {Ding}, \citenamefont {Huang},\ and\ \citenamefont
  {Ye}}]{chen2021compiling}%
  \BibitemOpen
  \bibfield  {author} {\bibinfo {author} {\bibfnamefont {J.}~\bibnamefont
  {Chen}}, \bibinfo {author} {\bibfnamefont {D.}~\bibnamefont {Ding}}, \bibinfo
  {author} {\bibfnamefont {C.}~\bibnamefont {Huang}}, \ and\ \bibinfo {author}
  {\bibfnamefont {Q.}~\bibnamefont {Ye}},\ }\href@noop {} {\bibfield  {journal}
  {\bibinfo  {journal} {Preprint at \url{https://arxiv.org/abs/2105.02398}}\ }
  (\bibinfo {year} {2021})}\BibitemShut {NoStop}%
\bibitem [{\citenamefont {Viebahn}\ \emph {et~al.}(2021)\citenamefont
  {Viebahn}, \citenamefont {Minguzzi}, \citenamefont {Sandholzer},
  \citenamefont {Walter}, \citenamefont {Sajnani}, \citenamefont {G\"org},\
  and\ \citenamefont {Esslinger}}]{viebahn2021suppressing}%
  \BibitemOpen
  \bibfield  {author} {\bibinfo {author} {\bibfnamefont {K.}~\bibnamefont
  {Viebahn}}, \bibinfo {author} {\bibfnamefont {J.}~\bibnamefont {Minguzzi}},
  \bibinfo {author} {\bibfnamefont {K.}~\bibnamefont {Sandholzer}}, \bibinfo
  {author} {\bibfnamefont {A.-S.}\ \bibnamefont {Walter}}, \bibinfo {author}
  {\bibfnamefont {M.}~\bibnamefont {Sajnani}}, \bibinfo {author} {\bibfnamefont
  {F.}~\bibnamefont {G\"org}}, \ and\ \bibinfo {author} {\bibfnamefont
  {T.}~\bibnamefont {Esslinger}},\ }\href {\doibase 10.1103/PhysRevX.11.011057}
  {\bibfield  {journal} {\bibinfo  {journal} {Phys. Rev. X}\ }\textbf {\bibinfo
  {volume} {11}},\ \bibinfo {pages} {011057} (\bibinfo {year}
  {2021})}\BibitemShut {NoStop}%
\bibitem [{\citenamefont {Bocharov}\ \emph {et~al.}(2017)\citenamefont
  {Bocharov}, \citenamefont {Roetteler},\ and\ \citenamefont
  {Svore}}]{bocharov2017factoring}%
  \BibitemOpen
  \bibfield  {author} {\bibinfo {author} {\bibfnamefont {A.}~\bibnamefont
  {Bocharov}}, \bibinfo {author} {\bibfnamefont {M.}~\bibnamefont {Roetteler}},
  \ and\ \bibinfo {author} {\bibfnamefont {K.~M.}\ \bibnamefont {Svore}},\
  }\href {\doibase 10.1103/PhysRevA.96.012306} {\bibfield  {journal} {\bibinfo
  {journal} {Phys. Rev. A}\ }\textbf {\bibinfo {volume} {96}},\ \bibinfo
  {pages} {012306} (\bibinfo {year} {2017})}\BibitemShut {NoStop}%
\bibitem [{\citenamefont {Gokhale}\ \emph {et~al.}(2019)\citenamefont
  {Gokhale}, \citenamefont {Baker}, \citenamefont {Duckering}, \citenamefont
  {Brown}, \citenamefont {Brown},\ and\ \citenamefont
  {Chong}}]{gokhale2019asymptotic}%
  \BibitemOpen
  \bibfield  {author} {\bibinfo {author} {\bibfnamefont {P.}~\bibnamefont
  {Gokhale}}, \bibinfo {author} {\bibfnamefont {J.~M.}\ \bibnamefont {Baker}},
  \bibinfo {author} {\bibfnamefont {C.}~\bibnamefont {Duckering}}, \bibinfo
  {author} {\bibfnamefont {N.~C.}\ \bibnamefont {Brown}}, \bibinfo {author}
  {\bibfnamefont {K.~R.}\ \bibnamefont {Brown}}, \ and\ \bibinfo {author}
  {\bibfnamefont {F.~T.}\ \bibnamefont {Chong}},\ }in\ \href {\doibase
  10.1145/3307650.3322253} {\emph {\bibinfo {booktitle} {Proceedings of the
  46th International Symposium on Computer Architecture}}}\ (\bibinfo
  {publisher} {{ACM}},\ \bibinfo {year} {2019})\ p.\ \bibinfo {pages}
  {554}\BibitemShut {NoStop}%
\bibitem [{\citenamefont {Campbell}(2014)}]{campbell2014enhanced}%
  \BibitemOpen
  \bibfield  {author} {\bibinfo {author} {\bibfnamefont {E.~T.}\ \bibnamefont
  {Campbell}},\ }\href {\doibase 10.1103/PhysRevLett.113.230501} {\bibfield
  {journal} {\bibinfo  {journal} {Phys. Rev. Lett.}\ }\textbf {\bibinfo
  {volume} {113}},\ \bibinfo {pages} {230501} (\bibinfo {year}
  {2014})}\BibitemShut {NoStop}%
\bibitem [{\citenamefont {Muralidharan}\ \emph {et~al.}(2017)\citenamefont
  {Muralidharan}, \citenamefont {Zou}, \citenamefont {Li}, \citenamefont
  {Wen},\ and\ \citenamefont {Jiang}}]{muralidharan2017overcoming}%
  \BibitemOpen
  \bibfield  {author} {\bibinfo {author} {\bibfnamefont {S.}~\bibnamefont
  {Muralidharan}}, \bibinfo {author} {\bibfnamefont {C.-L.}\ \bibnamefont
  {Zou}}, \bibinfo {author} {\bibfnamefont {L.}~\bibnamefont {Li}}, \bibinfo
  {author} {\bibfnamefont {J.}~\bibnamefont {Wen}}, \ and\ \bibinfo {author}
  {\bibfnamefont {L.}~\bibnamefont {Jiang}},\ }\href {\doibase
  10.1088/1367-2630/aa573a} {\bibfield  {journal} {\bibinfo  {journal} {New J.
  Phys.}\ }\textbf {\bibinfo {volume} {19}},\ \bibinfo {pages} {013026}
  (\bibinfo {year} {2017})}\BibitemShut {NoStop}%
\bibitem [{\citenamefont {Goss}\ \emph {et~al.}(2022)\citenamefont {Goss} \emph
  {et~al.}}]{goss2022high}%
  \BibitemOpen
  \bibfield  {author} {\bibinfo {author} {\bibfnamefont {N.}~\bibnamefont
  {Goss}} \emph {et~al.},\ }\href@noop {} {\bibfield  {journal} {\bibinfo
  {journal} {Preprint at \url{https://arxiv.org/abs/2206.07216}}\ } (\bibinfo
  {year} {2022})}\BibitemShut {NoStop}%
\bibitem [{\citenamefont {Shor}(1997)}]{shor1997polynomial}%
  \BibitemOpen
  \bibfield  {author} {\bibinfo {author} {\bibfnamefont {P.~W.}\ \bibnamefont
  {Shor}},\ }\href {\doibase 10.1137/s0097539795293172} {\bibfield  {journal}
  {\bibinfo  {journal} {SIAM J. Comput.}\ }\textbf {\bibinfo {volume} {26}},\
  \bibinfo {pages} {1484} (\bibinfo {year} {1997})}\BibitemShut {NoStop}%
\bibitem [{\citenamefont {Gidney}\ and\ \citenamefont
  {Eker{\aa{}}}(2021)}]{gidney2021how}%
  \BibitemOpen
  \bibfield  {author} {\bibinfo {author} {\bibfnamefont {C.}~\bibnamefont
  {Gidney}}\ and\ \bibinfo {author} {\bibfnamefont {M.}~\bibnamefont
  {Eker{\aa{}}}},\ }\href {\doibase 10.22331/q-2021-04-15-433} {\bibfield
  {journal} {\bibinfo  {journal} {{Quantum}}\ }\textbf {\bibinfo {volume}
  {5}},\ \bibinfo {pages} {433} (\bibinfo {year} {2021})}\BibitemShut {NoStop}%
\bibitem [{\citenamefont {Yoder}\ \emph {et~al.}(2016)\citenamefont {Yoder},
  \citenamefont {Takagi},\ and\ \citenamefont {Chuang}}]{yoder2016universal}%
  \BibitemOpen
  \bibfield  {author} {\bibinfo {author} {\bibfnamefont {T.~J.}\ \bibnamefont
  {Yoder}}, \bibinfo {author} {\bibfnamefont {R.}~\bibnamefont {Takagi}}, \
  and\ \bibinfo {author} {\bibfnamefont {I.~L.}\ \bibnamefont {Chuang}},\
  }\href {\doibase 10.1103/PhysRevX.6.031039} {\bibfield  {journal} {\bibinfo
  {journal} {Phys. Rev. X}\ }\textbf {\bibinfo {volume} {6}},\ \bibinfo {pages}
  {031039} (\bibinfo {year} {2016})}\BibitemShut {NoStop}%
\bibitem [{\citenamefont {Chao}\ and\ \citenamefont
  {Reichardt}(2018)}]{chao2018fault}%
  \BibitemOpen
  \bibfield  {author} {\bibinfo {author} {\bibfnamefont {R.}~\bibnamefont
  {Chao}}\ and\ \bibinfo {author} {\bibfnamefont {B.~W.}\ \bibnamefont
  {Reichardt}},\ }\href {\doibase 10.1038/s41534-018-0085-z} {\bibfield
  {journal} {\bibinfo  {journal} {npj Quantum Info.}\ }\textbf {\bibinfo
  {volume} {4}},\ \bibinfo {pages} {1} (\bibinfo {year} {2018})}\BibitemShut
  {NoStop}%
\bibitem [{\citenamefont {Nguyen}\ \emph {et~al.}(2022)\citenamefont {Nguyen}
  \emph {et~al.}}]{nguyen2022blueprint}%
  \BibitemOpen
  \bibfield  {author} {\bibinfo {author} {\bibfnamefont {L.~B.}\ \bibnamefont
  {Nguyen}} \emph {et~al.},\ }\href {\doibase 10.1103/PRXQuantum.3.037001}
  {\bibfield  {journal} {\bibinfo  {journal} {PRX Quantum}\ }\textbf {\bibinfo
  {volume} {3}},\ \bibinfo {pages} {037001} (\bibinfo {year}
  {2022})}\BibitemShut {NoStop}%
\bibitem [{\citenamefont {McKay}\ \emph {et~al.}(2017)\citenamefont {McKay},
  \citenamefont {Wood}, \citenamefont {Sheldon}, \citenamefont {Chow},\ and\
  \citenamefont {Gambetta}}]{mckay2017efficient}%
  \BibitemOpen
  \bibfield  {author} {\bibinfo {author} {\bibfnamefont {D.~C.}\ \bibnamefont
  {McKay}}, \bibinfo {author} {\bibfnamefont {C.~J.}\ \bibnamefont {Wood}},
  \bibinfo {author} {\bibfnamefont {S.}~\bibnamefont {Sheldon}}, \bibinfo
  {author} {\bibfnamefont {J.~M.}\ \bibnamefont {Chow}}, \ and\ \bibinfo
  {author} {\bibfnamefont {J.~M.}\ \bibnamefont {Gambetta}},\ }\href {\doibase
  10.1103/PhysRevA.96.022330} {\bibfield  {journal} {\bibinfo  {journal} {Phys.
  Rev. A}\ }\textbf {\bibinfo {volume} {96}},\ \bibinfo {pages} {022330}
  (\bibinfo {year} {2017})}\BibitemShut {NoStop}%
\bibitem [{\citenamefont {Abrams}\ \emph {et~al.}(2020)\citenamefont {Abrams},
  \citenamefont {Didier}, \citenamefont {Johnson}, \citenamefont {Silva},\ and\
  \citenamefont {Ryan}}]{abrams2020implementation}%
  \BibitemOpen
  \bibfield  {author} {\bibinfo {author} {\bibfnamefont {D.~M.}\ \bibnamefont
  {Abrams}}, \bibinfo {author} {\bibfnamefont {N.}~\bibnamefont {Didier}},
  \bibinfo {author} {\bibfnamefont {B.~R.}\ \bibnamefont {Johnson}}, \bibinfo
  {author} {\bibfnamefont {M.~P.~d.}\ \bibnamefont {Silva}}, \ and\ \bibinfo
  {author} {\bibfnamefont {C.~A.}\ \bibnamefont {Ryan}},\ }\href {\doibase
  10.1038/s41928-020-00498-1} {\bibfield  {journal} {\bibinfo  {journal} {Nat.
  Electron.}\ }\textbf {\bibinfo {volume} {3}},\ \bibinfo {pages} {744}
  (\bibinfo {year} {2020})}\BibitemShut {NoStop}%
\bibitem [{\citenamefont {Carignan-Dugas}\ \emph {et~al.}(2019)\citenamefont
  {Carignan-Dugas}, \citenamefont {Wallman},\ and\ \citenamefont
  {Emerson}}]{carignan2019bounding}%
  \BibitemOpen
  \bibfield  {author} {\bibinfo {author} {\bibfnamefont {A.}~\bibnamefont
  {Carignan-Dugas}}, \bibinfo {author} {\bibfnamefont {J.~J.}\ \bibnamefont
  {Wallman}}, \ and\ \bibinfo {author} {\bibfnamefont {J.}~\bibnamefont
  {Emerson}},\ }\href {\doibase 10.1088/1367-2630/ab1800} {\bibfield  {journal}
  {\bibinfo  {journal} {New J. Phys.}\ }\textbf {\bibinfo {volume} {21}},\
  \bibinfo {pages} {053016} (\bibinfo {year} {2019})}\BibitemShut {NoStop}%
\bibitem [{\citenamefont {Arute}\ \emph {et~al.}(2019)\citenamefont {Arute}
  \emph {et~al.}}]{arute}%
  \BibitemOpen
  \bibfield  {author} {\bibinfo {author} {\bibfnamefont {F.}~\bibnamefont
  {Arute}} \emph {et~al.},\ }\href {\doibase 10.1038/s41586-019-1666-5}
  {\bibfield  {journal} {\bibinfo  {journal} {Nature}\ }\textbf {\bibinfo
  {volume} {574}},\ \bibinfo {pages} {505} (\bibinfo {year}
  {2019})}\BibitemShut {NoStop}%
\bibitem [{\citenamefont {Flammia}\ and\ \citenamefont
  {Wallman}(2020)}]{flammia2020efficient}%
  \BibitemOpen
  \bibfield  {author} {\bibinfo {author} {\bibfnamefont {S.~T.}\ \bibnamefont
  {Flammia}}\ and\ \bibinfo {author} {\bibfnamefont {J.~J.}\ \bibnamefont
  {Wallman}},\ }\href {\doibase 10.1145/3408039} {\bibfield  {journal}
  {\bibinfo  {journal} {ACM Trans. Quantum Comput.}\ }\textbf {\bibinfo
  {volume} {1}},\ \bibinfo {pages} {1} (\bibinfo {year} {2020})}\BibitemShut
  {NoStop}%
\bibitem [{\citenamefont {Hashim}\ \emph {et~al.}(2021)\citenamefont {Hashim}
  \emph {et~al.}}]{hashim2021randomized}%
  \BibitemOpen
  \bibfield  {author} {\bibinfo {author} {\bibfnamefont {A.}~\bibnamefont
  {Hashim}} \emph {et~al.},\ }\href {\doibase 10.1103/PhysRevX.11.041039}
  {\bibfield  {journal} {\bibinfo  {journal} {Phys. Rev. X}\ }\textbf {\bibinfo
  {volume} {11}},\ \bibinfo {pages} {041039} (\bibinfo {year}
  {2021})}\BibitemShut {NoStop}%
\end{thebibliography}%

\clearpage
\section*{Methods}
\renewcommand{\figurename}{Extended Data Fig.\!}
\renewcommand{\tablename}{Extended Data Table\!}

\setcounter{figure}{0}

\noindent\textbf{Experimental setup}

    \begin{table}[b]
    \begin{tabular}{ | m{10em} | m{6.5em} m{8.5em} | } 
     \hline
     Component & Brand & Model \\
     \hline
     Dilution fridge  & BlueFors  & XLD  \\
     Control chassis  & Keysight  & PXI M9023A  \\
     AWG  & Keysight  & PXI M3202A  \\
     Digitizer  & AlazarTech  & ATS9373  \\
     LO RF source  & Keysight  & MXG N5183B  \\
     Spectrum analyzer & Keysight & N9320B \\
     Frequency standard & SRS & FS725 \\
     Para. amplifier  & MIT LL  & TWPA  \\
     HEMT  & LNF  & LNC4$\_$8C  \\
     TWPA pump  & Hittite  & HMC M2100  \\
     IQ mixer  & Marki  & MLIQ-0416  \\
     Bias-Tee & Mini-Circuits & ZX85-12G-S+ \\
     Attenuator & XMA & 4882-6240\\
     IR filter mixture & Laird & Eccosorb\textsuperscript{\tiny\textregistered}CR-110\\
    \hline
    \end{tabular}
    \caption{Component brands and models used in the experimental setup.}
    \label{tab:components}
    \end{table}
    
    \begin{figure*}
        \includegraphics[width=\textwidth]{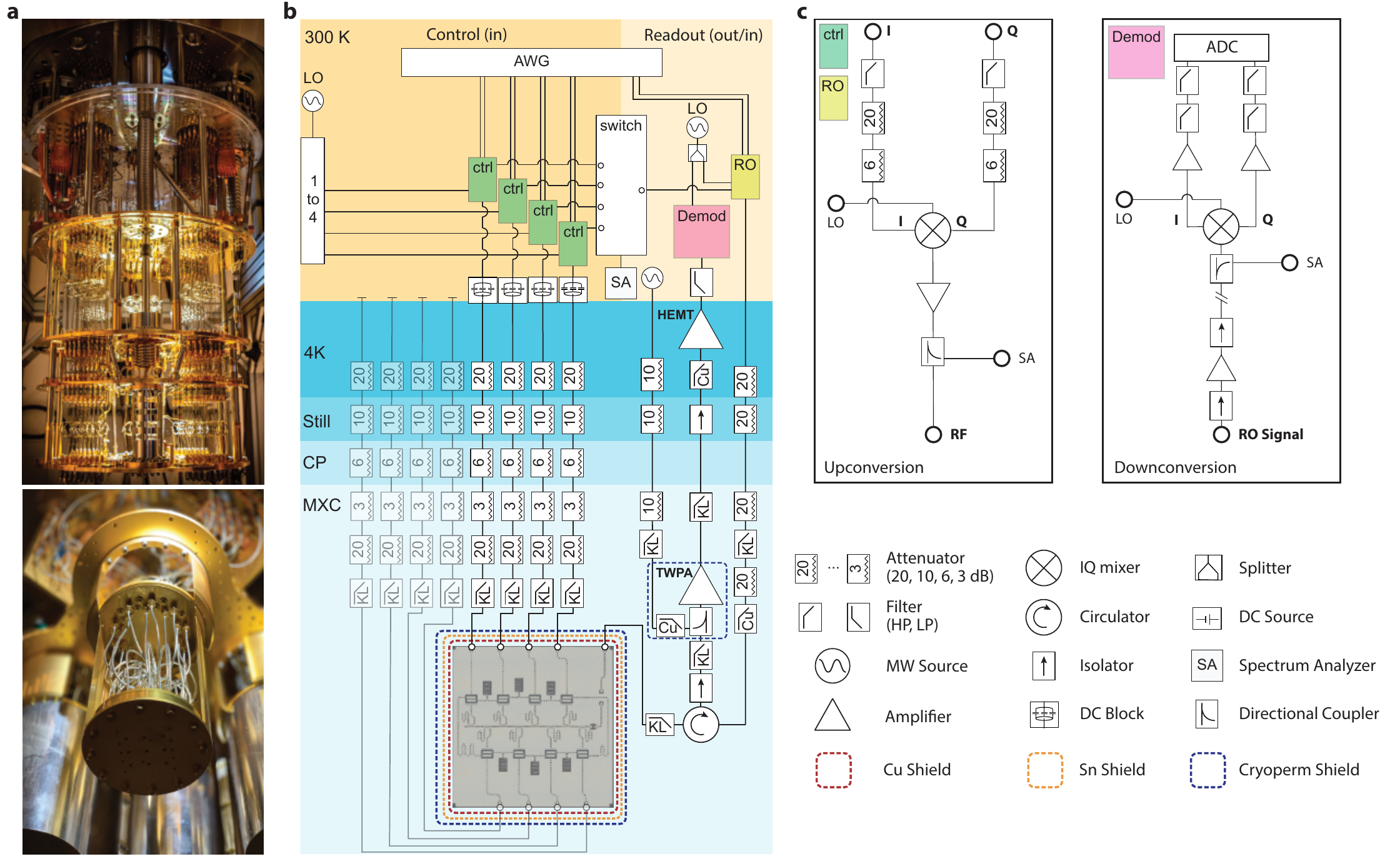}
        \caption{\label{figs1}\textbf{Experimental setup}. \textbf{a,} Dilution unit setup. \textbf{b,} Cryogenic wiring diagram. \textbf{c,} Room temperature control and readout circuitry.}
    \end{figure*}
    
    \noindent Extended Data Figure~\ref{figs1} shows the cryogenic setup and electronic wiring diagrams. The superconducting quantum processor used to perform the experiment is housed in a dilution refrigerator and operates at approximately 12 mK. The wiring is separated into two sides: the input side with predominantly attenuators, and the output side with mainly circulators and amplifiers. A combination of commercial $\mathrm{K\&L}$ and in-house infrared Eccosorb\textsuperscript{\tiny\textregistered} low-pass filters are added on both sides to mitigate high-frequency noise. The sample box is protected by tri-layer shields, with the copper shield further painted with \textit{Berkeley black} mixture consisting of STYCAST 2850FT, silica, and carbon powder. The enclosure is made light-tight using indium seals. The parametric amplifier is placed inside a separate magnetic shield, and its pump line is also connected to low-pass filters to ensure suppression of high-frequency noise. 
    
    Due to resource constraint, only four qubits are connected to external circuitry during the experiment. All the input and output signals are diverted via directional couplers and power splitters to a spectrum analyzer for calibration. The pulses used to control the qubits are generated by an arbitrary waveform generator (AWG) at 1 GSa/s and upconverted using IQ mixers in combination with a local oscillator (LO) carrier tone. DC offsets using bias tees and local phases of the signals are calibrated to null carrier leakage. These offset parameters are found numerically by minimizing the signals at the carrier frequency using active feedback via the spectrum analyzer (SA) and COBYLA optimization.
    
    The input signals are attenuated, filtered, and DC-blocked at room temperature to reduce noise inherent to electrical components inside the AWG. 26 dB of additional attenuation is added at the outputs of the AWG to fully utilize its dynamic range. The readout pulses are upconverted in a similar fashion. The upconversion circuit wiring is shown in the left panel of Extended Data Fig.~\ref{figs1}\textbf{c}. While the qubits are manipulated via individual on-chip coplanar traces, the readout is performed in multiplexed fashion via a common bus. The microwave tones going into the fridge are checked using the SA for consistency.
    
    The signal reflected from a readout resonator is circulated and amplified by a traveling wave parametric amplifier (TWPA) at the mixing chamber (MXC) plate, then by a high-electron-mobility-transistor (HEMT) amplifier at the 4-K plate, and finally by a room temperature amplifier before reaching the downconversion circuit. Here, an IQ mixer combines the same LO tone with the outgoing signal to produce an IF pulse that carries the information about the qubit. The measurement signal is further amplified at room temperature and goes through low-pass filters to suppress additional noise coming from the amplification. Finally, it is digitized at a rate of 1 GSa/s by an analog-to-digital converter (ADC) board attached directly to the PXI slot of the acquisition computer and demodulated via software. The downconversion circuit is shown in the right panel of Extended Data Fig.~\ref{figs1}\textbf{c}. All electronic instruments are synchronized using a rubidium atomic clock. The components used to construct the experiment are listed in Extended Data Table \ref{tab:components}.
\\

\noindent\textbf{Device tuneup and characterization}

    \noindent The device used in this experiment consists of eight single-junction transmon qubits. Each is formed by two superconducting electrodes sandwiching a thin layer of aluminum oxide, resulting in a Josephson junction with Josephson energy $E_J$ shunted by a capacitor with charging energy $E_C$. These characteristic energies define the spectrum of the qubit, resembling that of an anharmonic oscillator with transition frequency between the first two levels $\omega_{01}/2\pi\approx \sqrt{8E_JE_C}-E_C$ and anharmonicity $\alpha\approx -E_C$ \cite{koch2007charge}. They are pairwise-coupled to mutual  coplanar-waveguide (CPW) resonators (Fig.~\ref{fig1}\textbf{a}), resulting in an effective capacitive coupling~\cite{majer2007coupling}. These couplers are designed to have frequencies at around 7 GHz and a resonator-qubit coupling strength of approximately 70 MHz. Each qubit is coupled to a separate control line and a CPW readout resonator. All readout resonators are coupled to a common bus which also serves as a Purcell filter.
        
\begin{figure*}[t]
        \includegraphics[width=\textwidth]{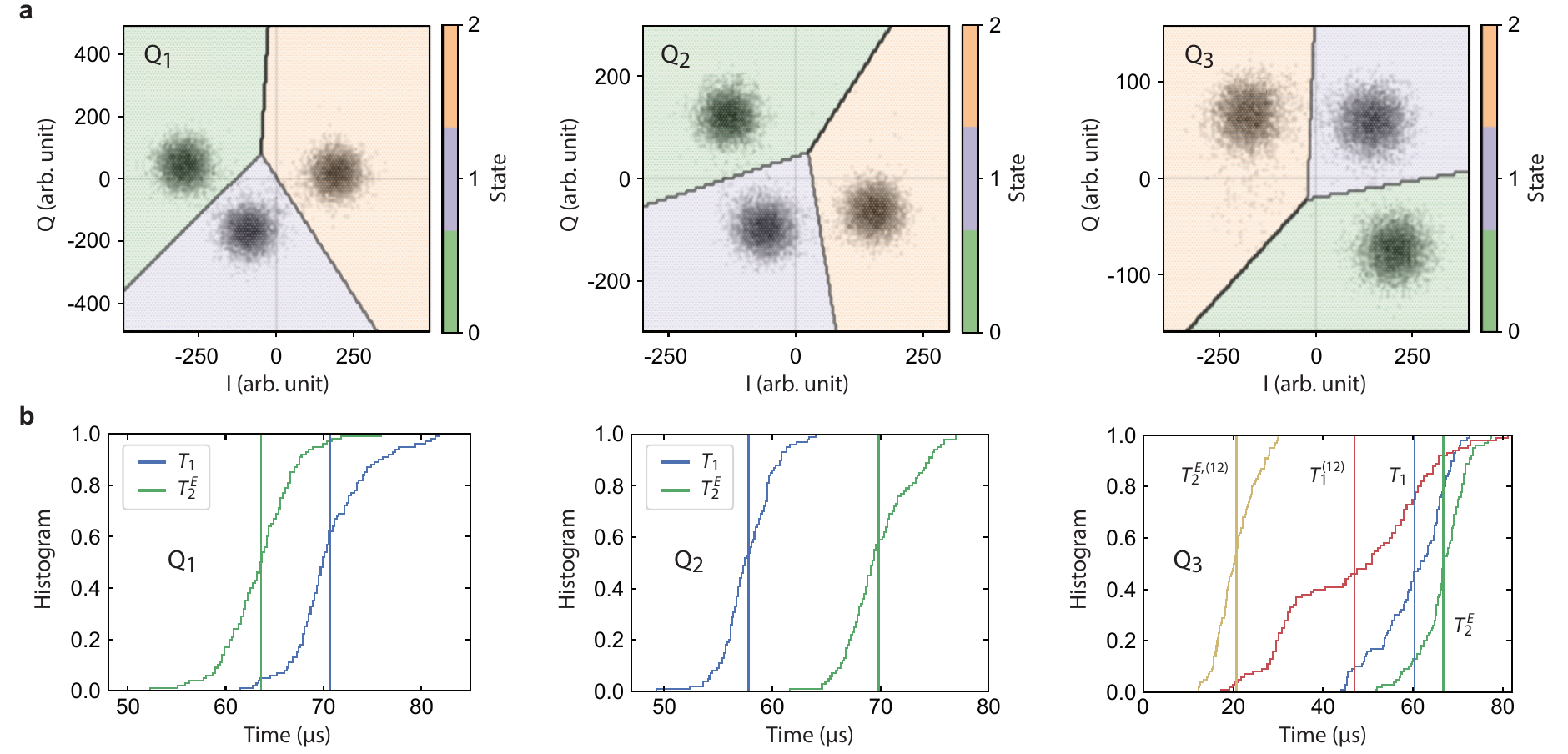}
        \caption{\label{figs2}\textbf{Device characterization}. \textbf{a,} Dispersive readout histogram showing the IQ signal of the resonators corresponding to the qubits in state $|0\rangle$, $|1\rangle$, and $|2\rangle$. \textbf{b,}. Coherence statistics acquired continuously for 100 iterations. The vertical lines indicate their average values.}
    \end{figure*}

    The extracted parameters of the qubits and their readout used in this experiment are listed in Extended Data Table \ref{tab:qparameters}. The readout and qubit frequencies are measured using microwave spectroscopy. The TWPA pump tone is calibrated using a vector network analyzer to optimize the signal-to-noise-ratio. The readout pulse is set to be 1-$\mathrm{\mu s}$ long, and its amplitude is adjusted to optimize the measurement fidelity $\mathcal{F}_\text{RO}(|i\rangle)=P(i|i)$, where $P(x|y)$ is the probability that the qubit initialized in state $|y\rangle$ is measured to be in state $|x\rangle$. The qubit-state-dependent readout signals are shown on the IQ planes in Extended Data Fig.~\ref{figs2}\textbf{a}.
    
    The relaxation time $T_1$ and echo dephasing time $T_2^E$ are extracted by applying a pulse sequence with variable length to the qubit, measuring the signal from the readout resonators, then fitting the data to appropriate functions with a decay parameter. Typical coherence time statistics consisting of ensembles of 100 individual measurements are shown in Extended Data Fig.~\ref{figs2}\textbf{b}. We also measure the coherence statistics of the $|1\rangle \leftrightarrow |2\rangle$ transition for Q$_3$, with $\overline{T}_1^{(12)}=46.96~\mathrm{\mu s}$ and $\overline{T}_2^{E,(12)}=20.97~\mathrm{\mu s}$.
    
    
    Single-qubit pulses have a Gaussian shape and are set to be 30-$\mathrm{ns}$ long. A series of Ramsey sequences is applied to extract the correct qubit frequencies. Their amplitudes are calibrated for $\text{R}_\text{X}(\pi/2)$ pulses by repeatedly applying even numbers of them up to $n_\mathrm{pulse}=200$ and measuring the outcomes to zero-in the correct parameters. By using the $\text{R}_\text{X}(\pi/2)$ gate in combination with virtual-$\mathrm{Z}$ gates, single qubit rotations can be implemented using the $\mathrm{ZXZXZ}$ decomposition~\cite{mckay2017efficient}, which significantly reduces the calibration time and complexity. Single-qubit gate fidelities extracted via streamline randomized benchmarking are included in Extended Data Table~\ref{tab:qparameters}.
    
    \begin{table}[b]
        \begin{tabular}{ | m{7em} | m{5em} m{5em} m{5em} | } 
         \hline
          & Q$_1$ & Q$_2$ & Q$_3$ \\
         \hline
         \hline
         $\omega_{01}/2\pi$ (GHz) & 5.23 & 5.32 & 5.44\\
         $\alpha$ (GHz) & -0.26 & -0.27 & -0.27\\
         $\omega_\mathrm{RO}/2\pi$ (GHz) & 6.23 & 6.32 & 6.44\\
         $\overline{T}_1~(\mathrm{\mu s})$ & 70.67 & 57.81 & 60.35\\
         $\overline{T}_2^E~(\mathrm{\mu s})$ & 63.62 & 69.87 & 66.77\\
         $\mathcal{F}_\text{RO}$ ($|0\rangle$) & 99.6\% & 99.3\% & 99.1\%\\
         $\mathcal{F}_\text{RO}$ ($|1\rangle$)& 98.2\% & 97.4\% & 97.5\%\\
         $\mathcal{F}_\text{RO}$ ($|2\rangle$)& 96.8\% & 96.8\% & 95.8\%\\
         $\mathcal{F}_\text{1Q}$ (isolated) & 99.88(1)\% & 99.85(1)\% & 99.80(1)\%\\
         $\mathcal{F}_\text{1Q}$ (joint) & 99.68(1)\% & 99.4(1)\% & 99.72(3)\%\\
        \hline
        \end{tabular}
        \caption{Relevant device parameters.}
        \label{tab:qparameters}
        \end{table}
        
    Classical microwave crosstalk is present in the device and degrades its performance via two distinct processes. Firstly, an off-resonant microwave tone applied to the qubit slightly dresses its resonant frequency as discussed in the main text, leading to phase errors during execution of intended quantum circuits. Secondly, a leaked microwave tone applied to one qubit at the frequency of another qubit coupled to it induces a $\mathrm{ZX}$ interaction, which is commonly known as the cross-resonance effect, leading to spurious entanglement between the qubits. Thus, it is important to detect and suppress such crosstalk to improve chip-scale performance. To accomplish this, additional microwave tones are applied simultaneously to neighboring qubits during the gate operation of the intended qubits. Their amplitudes and phases are tuned up to destructively interfere with the crosstalk tones. The parameters are further optimized using covariance matrix adaptation (CMA) with simultaneous randomized benchmarking. Typically, the simulataneous single-qubit gate errors are considerably higher than the isolated errors (Extended Data Table~\ref{tab:qparameters}). Although we routinely observe high isolated single-qubit gate fidelities immediately following a calibration round, the simultaneous fidelities tend to saturate toward the reported numbers, which are more relevant for the present experiment.
    
    Using optimal readout and single-qubit gate parameters, a readout confusion matrix is extracted for each qubit by preparing it in a certain state and measuring the probability for it to be in $|0\rangle$, $|1\rangle$, or $|2\rangle$ states. This results in a matrix that is ideally diagonal with entries equal to 1. Thus, a correction matrix can be found by inverting the confusion matrix. The correction matrix is subsequently applied to all measurement outcomes to compensate for readout errors.
    \\

    \begin{figure}[t]
            \includegraphics[width=0.5\textwidth]{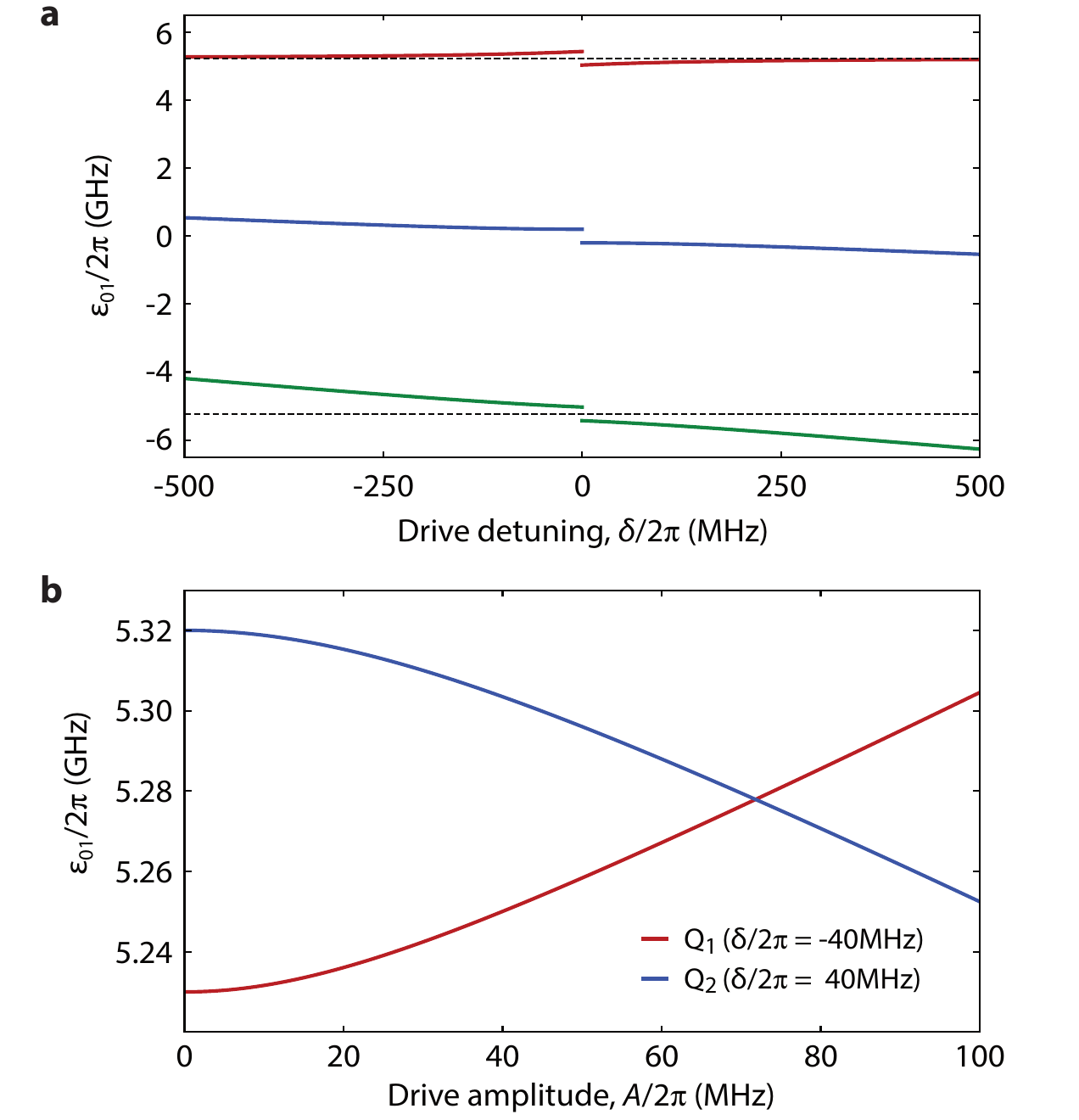}
            \caption{\label{figsf}\textbf{Floquet quasienergies}. \textbf{a,} Transition energies between $|u_0(t)\rangle_\text{F}$ and $|u_1(t)\rangle_\text{F}$ with drive amplitude 200 MHz, qubit frequency 5.23 GHz, and varying drive detuning $\delta$. The dashed lines represent the qubit frequency.  
            \textbf{b,} Dependence of $\varepsilon_{01}$ on the drive amplitude. By driving qubit Q$_1$(Q$_2$) with red(blue) detuned microwave, two qubits can be brought into resonance.}
        \end{figure}

\noindent\textbf{Floquet states and quasienergies}

\noindent We consider a periodically driven qubit system described by the Hamiltonian 
\begin{equation}
\hat{\mathcal{H}}_\text{q}(t)/\hbar=-\frac{\omega_\text{q}}{2}\hat{\sigma}_\text{z}+ A\ \!\text{cos}(\omega_\text{d}t+\varphi)\hat{\sigma}_\text{x}.
\end{equation}
For a Hamiltonian with a period of $T=2\pi/\omega_\text{d}$, we can find a time-periodic Floquet state, $|u(t)\rangle_\text{F}=|u(t+T)\rangle_\text{F}$, with quasienergy $\hbar\varepsilon$ through the Floquet equation,
\begin{equation}
\left(\hat{\mathcal{H}}_\text{q}(t)-i\hbar\partial_t\right)|u(t)\rangle_\text{F}=\hbar\varepsilon|u(t)\rangle_\text{F}.
\label{Method:Floquet_eq}
\end{equation}
Note that this equation is connected to the Schr\"{o}dinger equation via the relation  $|u(t)\rangle_\text{F}=e^{i\varepsilon t}|\psi(t)\rangle$. To find the solutions of Eq.~(\ref{Method:Floquet_eq}), we write the Floquet state as a Fourier series comprising of time-independent states $|u^k\rangle$, where $k$ is an integer,
\begin{equation}
|u(t)\rangle_\text{F}=\sum_k e^{ik\omega_\text{d}t}|u^k\rangle,
\label{Fourier}
\end{equation}
and then inserting it into the Floquet equation. By grouping terms which have the same Fourier frequencies, we can obtain
$\varepsilon|u^k\rangle=(-\frac{\omega_\text{q}}{2}\hat{\sigma}_\text{z}+k\omega_\text{d})|u^k\rangle+  e^{i\varphi}\frac{A}{2}\hat{\sigma}_\text{x}|u^{k-1}\rangle +e^{-i\varphi}\frac{A}{2}\hat{\sigma}_\text{x}|u^{k+1}\rangle$. Under the conditions $A\ll\omega_\text{q},\omega_\text{d}$ and $\omega_\text{q}\leq 2\omega_\text{d}$, this can be put into the matrix form,
\begin{equation}
\begin{bmatrix}
\text{-}\frac{\omega_\text{q}}{2}\hat{\sigma}_\text{z} +k\omega_\text{d} & e^{i\varphi}\frac{A}{2}\hat{\sigma}_\text{x} \\
e^{-i\varphi}\frac{A}{2} \hat{\sigma}_\text{x} & \text{-}\frac{\omega_\text{q}}{2}\hat{\sigma}_\text{z}+(k-1)\omega_\text{d}
\end{bmatrix}\!
\begin{bmatrix}
|u^k\rangle \\
|u^{k\text{-}1}\rangle
\end{bmatrix}=\varepsilon\begin{bmatrix}
|u^k\rangle \\
|u^{k\text{-}1}\rangle
\end{bmatrix}.
\end{equation}
Its eigenstates yield the Floquet basis states using Eq.~(\ref{Fourier}), 
\begin{eqnarray}
&|u_+(t)\rangle_\text{F}\propto \left(\delta+\sqrt{A^2+\delta^2}\right)|0\rangle+e^{-i(\omega_\text{d} t+\varphi)}A|1\rangle, \nonumber\\
&|u_-(t)\rangle_\text{F}\propto \left(\delta-\sqrt{A^2+\delta^2}\right)|0\rangle+e^{-i(\omega_\text{d} t+\varphi)}A|1\rangle,
\label{Fstate}
\end{eqnarray}
and its eigenvalues give the quasienergies of $e^{ik\omega_\text{d}t}|u_{\pm}(t)\rangle_\text{F}$ as
\begin{equation}
    \varepsilon_\pm^k = \left(k-\frac{1}{2}\right)\omega_\text{d}\pm\frac{1}{2} \sqrt{A^2+\delta^2},
    \label{quasienergy}
\end{equation}
where $\delta=\omega_\text{d}-\omega_\text{q}$ is the drive detuning. In this work, we allocate the index of the Floquet basis to match with that of the bare basis, i.e., $|\langle n|u_n(t)\rangle_\text{F}|^2>0.5$. Specifically, if $\delta>0$, we define $|u_{+(-)}(t)\rangle_\text{F}$ as $|u_{0(1)}(t)\rangle_\text{F}$ and $\varepsilon_{+(-)}$ as $\varepsilon_{0(1)}$, otherwise the index is inverted. Thus, the quasienergy difference, $\varepsilon_{01}\equiv\varepsilon_{1}-\varepsilon_{0}$, abruptly changes at $\delta=0$ (Extended Data Fig.~\ref{figsf}\textbf{a}).\\

 
\noindent\textbf{Heisenberg interactions between Floquet states}

\noindent As described by Eq.~(\ref{eqn:floquet_heisenberg}), the interaction strengths of the Floquet-engineered Heisenberg Hamiltonian are given as  
\begin{eqnarray}
&&J_\text{XY}=J\langle c_{01}^{(1)}c_{10}^{(2)}\rangle_t=J\langle c_{10}^{(1)}c_{01}^{(2)}\rangle_t,\nonumber\\
&&J_\text{ZZ}=J\langle
c_{11}^{(1)}c_{11}^{(2)}+c_{00}^{(1)}c_{00}^{(2)}-c_{00}^{(1)}c_{11}^{(2)}-c_{11}^{(1)}c_{00}^{(2)}\rangle_t,
\end{eqnarray}
where $c_{ab}^{(n)}=\langle\psi_a^{(n)}(t)|\hat{\sigma}_\text{x}^{(n)}|\psi_b^{(n)}(t)\rangle$ is for qubit Q$_n$ and $\langle...\rangle_t$ denotes the time-average value. 
For the transverse (XY) spin-exchange interaction, we drive Q$_1$(Q$_2$) with red(blue) detuned microwaves to bring their quasienergy differences $\varepsilon^{(n)}_{01}$ into resonance. Specifically, this condition is satisfied for quasienergy transitions $e^{ik\omega_d^{(n)}t}|u_0^{(n)}(t)\rangle_\text{F}\leftrightarrow e^{i(k+1)\omega_d^{(n)}t}|u_1^{(n)}(t)\rangle_\text{F}$ when mapped by a certain drive amplitude (Eq.~(\ref{quasienergy}) and Extended Data Fig.~\ref{figsf}\textbf{b}). Accordingly, the Floquet states' coefficients read
\begin{eqnarray}
c^{(1)}_{01}&&=\langle\psi^{(1)}_0(t)|\hat{\sigma}_\text{x}^{(1)}|\psi^{(1)}_1(t)\rangle \nonumber\\
&&=e^{-i\varepsilon^{(1)}_{01}t}{}\langle u^{(1)}_-(t)|_\text{F}\hat{\sigma}_\text{x}^{(1)}e^{i\omega^{(1)}_\text{d}t}|u^{(1)}_+(t)\rangle_\text{F}
\end{eqnarray}
and
\begin{eqnarray}
c^{(2)}_{10}&&=\langle\psi^{(2)}_1(t)|\hat{\sigma}_\text{x}^{(2)}|\psi^{(2)}_0(t)\rangle \nonumber\\
&&=e^{i\varepsilon^{(2)}_{01}t}{}\langle u^{(2)}_-(t)|_\text{F}e^{-i\omega^{(2)}_\text{d}t}\hat{\sigma}_\text{x}^{(2)}|u^{(2)}_+(t)\rangle_\text{F}.
\end{eqnarray}
When the blue- and red-detuned drives have the same detuning $|\delta|$ and amplitude $A$,  the time-average XY interaction strength is obtained as 
\begin{equation}
|J_\text{XY}|=J|\langle c^{(1)}_{01} c^{(2)}_{10}\rangle_t|=J\frac{(|\delta|+\sqrt{A^2+\delta^2})^2}{4(A^2+\delta^2)},
\end{equation}
using Eq.~(\ref{Fstate}). On the other hand, the relevant coefficients for the longitudinal (ZZ) spin-spin interaction read
\begin{eqnarray}
c^{(n)}_{\pm\pm}&&=\langle u^{(n)}_\pm(t)|_\text{F}\hat{\sigma}_\text{x}^{(n)}|u^{(n)}_\pm(t)\rangle_\text{F}\nonumber\\
&&=\frac{A_n(\delta_n\pm\sqrt{A_n^2+ \delta_n^2} )\cos\left(\omega_{\text{d}}^{(n)}t+\varphi_n\right)}{A_n^2+\delta_k\left(\delta_n\pm\sqrt{A_n^2+ \delta_n^2}\right)}.
\end{eqnarray}
When the microwave drives for Q$_1$ and Q$_2$ are both detuned to red or blue,
\begin{eqnarray}
&&c^{(1)}_{11}c^{(2)}_{11}+c^{(1)}_{00}c^{(2)}_{00}-c^{(1)}_{11}c^{(2)}_{00}-c^{(1)}_{00}c^{(2)}_{11} \nonumber\\
&&= c^{(1)}_{++}c^{(2)}_{++}+c^{(1)}_{--}c^{(2)}_{--}-c^{(1)}_{++}c^{(2)}_{--}-c^{(1)}_{--}c^{(2)}_{++} \nonumber\\
&&=\frac{4A_1A_2 \cos(\omega_\text{d}^{(1)}t+\varphi_1)\cos(\omega_\text{d}^{(2)}t+\varphi_2)}{\sqrt{(A_1^2+\delta_1^2)(A_2^2+\delta_2^2)}}.
\end{eqnarray}
For $\omega_\text{d}^{(1)}=\omega_\text{d}^{(2)}$, the time-averaged ZZ interaction strength is thus given as 
\begin{eqnarray}
J_\text{ZZ} &&=J\langle
c_{11}^{(1)}c_{11}^{(2)}+c_{00}^{(1)}c_{00}^{(2)}-c_{00}^{(1)}c_{11}^{(2)}-c_{11}^{(1)}c_{00}^{(2)}\rangle_t\nonumber \\ &&=J\frac{2A_1A_2\cos(\varphi_1-\varphi_2)}{\sqrt{(A_1^2+\delta_1^2)(A_2^2+\delta_2^2)}}.
\end{eqnarray}\\

\noindent\textbf{ZZ interaction measurements} 
 \begin{figure*}[t]
        \includegraphics[width=\textwidth]{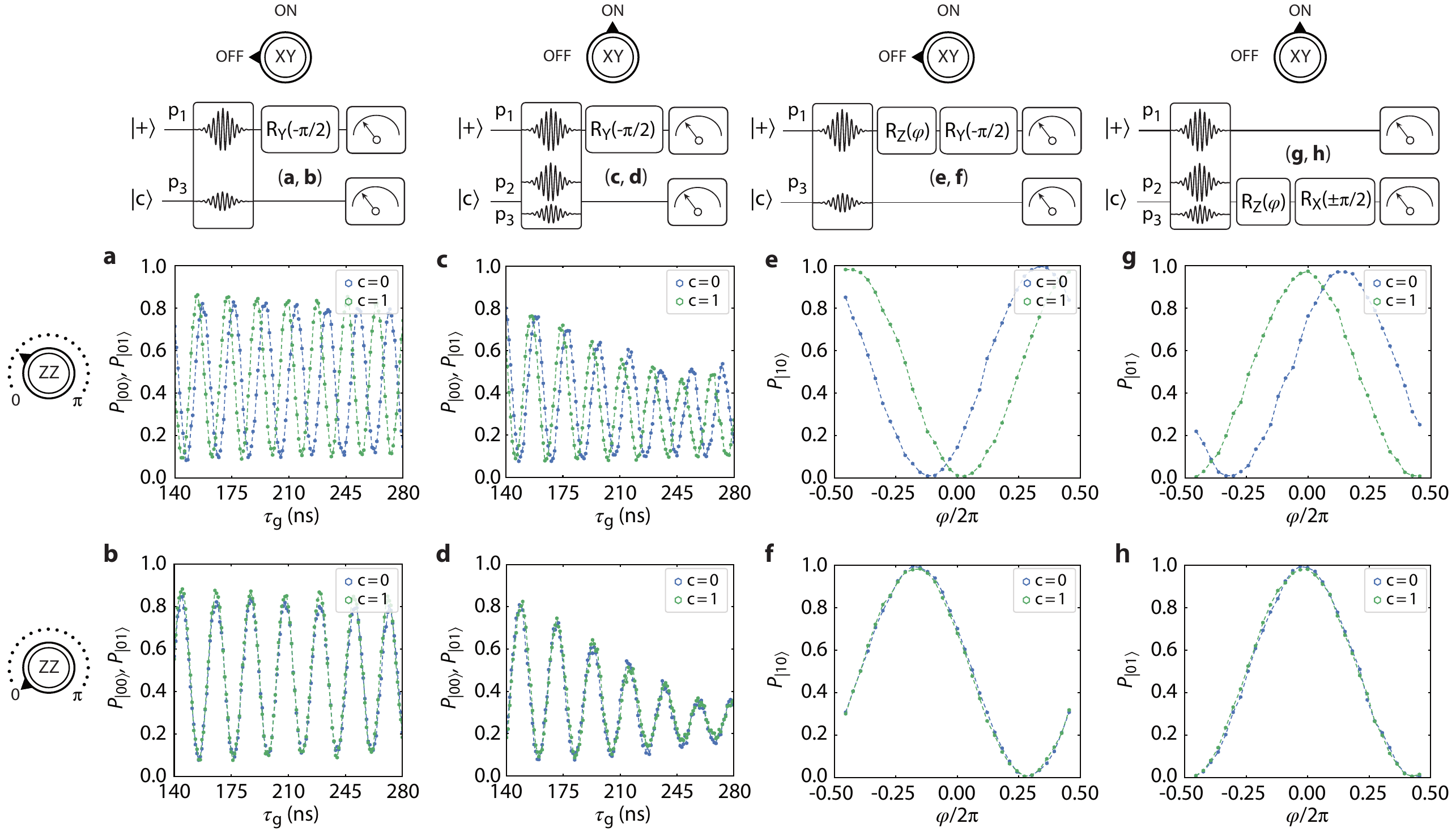}
        \caption{\label{figs3}\textbf{Spin-spin interaction measurement}. \textbf{a}-\textbf{d,} Dependence of Q$_1$'s $\mathrm{Z}$ rate on Q$_2$'s state, with transverse coupling off (\textbf{a}-\textbf{b}) and on (\textbf{c}-\textbf{d}). \textbf{e}-\textbf{f,} Dependence of Q$_1$'s $\mathrm{Z}$ phase on Q$_2$'s state, without transverse coupling. The phase difference results from $\Phi_\mathrm{ZZ}$. \textbf{g}-\textbf{h,} When the two-qubit states $|10\rangle$ and $|01\rangle$ are swapped via $\mathrm{XY}$ coupling, the sequence instead includes the projection gate on Q$_2$. The angle polarity is chosen according to the initial control state. The analysis takes into account the $2\pi$ phase difference between the curves when $\Phi_\mathrm{ZZ}=0$ to mimic the result without transverse coupling.}
    \end{figure*}
    
    \noindent The $\mathrm{ZZ}$ rate can be estimated using a time-efficient and reliable Ramsey-like experiment. The sequence consists of initializing one qubit in a superposition state, then applying the intended pulses and measuring it along the $\mathrm{Z}$ axis. The $\mathrm{Z}$ rotation resulting from the frame frequency difference between the bare qubit and the Floquet qubit manifests as an oscillation of the qubit's population with respect to the pulse duration. For a finite spin-spin coupling, the two oscillation frequencies corresponding to the other qubit in the ground and excited states are subtracted to find the $\mathrm{ZZ}$ rate. As this always works for both zero and finite $\mathrm{XY}$ coupling (Extended Data Fig.~\ref{figs3}\textbf{a}-\textbf{d}), it is a versatile method to approximate the longitudinal coupling. However, its accuracy  is limited by the fast oscillations and subsequently the fitting errors. 
    
    To accurately measure the $\mathrm{ZZ}$ angle at the end of a pulse with a fixed duration $\tau_\text{g}$, a sequence with the addition of a single-qubit $\mathrm{Z}$ gate with a variable phase $\varphi$ is used. The measured qubit population depends on $\varphi$, and $\Phi_\mathrm{ZZ}$ is the phase difference corresponding to the other qubit being in $|0\rangle$ and $|1\rangle$ states (Extended Data Fig.~\ref{figs3}\textbf{e}-\textbf{f}). The measurement also works when there is a complete population swap between $|10\rangle$ and $|01\rangle$. In this case, the $\mathrm{Z}$ gate is applied to the other qubit instead, with the projection gate's polarity depending on its initial state. $\Phi_\mathrm{ZZ}=0$ then corresponds to the phase difference of $2\pi$, which can be subtracted from the results to give overlapping trajectories for zero $\mathrm{ZZ}$ coupling. We note that since the qubit is expected to be in the ground state if there is no phase gate induced by the pulse, a similar approach is used to calibrate single-qubit $\mathrm{Z}$ gates. Finally, the method can be extended to analyze a three-qubit $\mathrm{CCZ}$ entanglement.
    
    For simultaneous transverse coupling resulting in arbitrary swapping angle, we find tomographic reconstruction to be the most reliable. In this approach, quantum state tomography and quantum process tomography are performed to find the state or process matrices resulting from the applied gate. An optimization routine is then utilized to extract the swapping (XY) and ZZ angles. The accuracy of this method is inherently limited by the SPAM errors of the tomography procedures. 
    \\

\noindent\textbf{Two-qubit gate calibration}\\ 
    \noindent The programmable interactions can be employed to implement two-qubit gates in a straightforward manner. Since the $\mathrm{XY}$ and $\mathrm{ZZ}$ operations commute, we can decompose the Heisenberg unitary as
    $\hat{U}_\mathrm{XXZ}(\Theta_\text{XY},\Phi_\text{ZZ}) =e^{-i\hat{\mathcal{H}}_\text{XXZ}t/\hbar} = \hat{U}_\mathrm{XY}(\Theta_\text{XY})\times \hat{U}_\mathrm{ZZ}(\Phi_\text{ZZ})$, where
    \begin{equation}
    \begin{split}
        \hat{U}_\mathrm{XY}(\Theta_\text{XY})&=\,\text{exp}\left[ -i\frac{\Theta_\text{XY}}{2}(\hat{X}\hat{X}+\hat{Y}\hat{Y}) \right] \\
        &= \begin{pmatrix}
                    1 & 0 & 0 & 0\\
                    0 & \cos\left(\Theta_\text{XY} \right) & -i\sin\left(\Theta_\text{XY} \right) & 0\\
                    0 & -i\sin\left(\Theta_\text{XY} \right) & \cos\left(\Theta_\text{XY} \right) & 0\\
                    0 & 0 & 0 & 1
                \end{pmatrix}
    \end{split}
    \end{equation}
    and
    \begin{equation}
    \begin{split}
        \hat{U}_\mathrm{ZZ}(\Phi_\text{ZZ})&= \,\text{exp}\left[ -i\frac{\Phi_\text{ZZ}}{2}\hat{Z}\hat{Z} \right] \\
        &= \begin{pmatrix}
                    e^{-i\Phi_\text{ZZ}/2} & 0 & 0 & 0\\
                    0 & e^{i\Phi_\text{ZZ}/2} & 0 & 0\\
                    0 & 0 & e^{i\Phi_\text{ZZ}/2} & 0\\
                    0 & 0 & 0 & e^{-i\Phi_\text{ZZ}/2}
                \end{pmatrix}.
    \end{split}
    \end{equation}
    
    Naturally, the coherent flip-flop between $|10\rangle$ and $|01\rangle$ in Fig.~\ref{fig2}\textbf{b} is equivalent to the rotation of the transverse coupling angle $\Theta_\text{XY}$, with $\Theta_\text{XY}=\pi/2$ corresponding to a full swap (the oscillation angle is $2\Theta_\text{XY}$). The measured longitudinal coupling angle $\Phi_\text{ZZ}$ in Fig.~\ref{fig2}\textbf{c} illustrates a control-phase entangling operation up to local Z rotations, with $\Phi_\text{ZZ}=\pi/2$ corresponding to a $\mathrm{CZ}$ gate following the relation $\hat{U}_\mathrm{CZ} = \exp\left[-i\pi(\hat{Z}\hat{Z} + \hat{I}\hat{I} - \hat{Z}\hat{I} - \hat{I}\hat{Z})/4 \right]$. Therefore, two-qubit gates can be calibrated up to local $\mathrm{Z}$ gates using the observed interactions. Notably, the realized $\mathrm{iSWAP}$ and $\mathrm{SWAP}$ gates arise directly from the $\mathrm{XX}$ and isotropic $\mathrm{XXX}$ Heisenberg models, while the $\mathrm{CZ}$ gate results from the pure spin-spin coupling of the transverse-field Ising model. In sum, the realized Floquet system can be used as a robust quantum many-body simulator with high-fidelity gates as consequences.
    
    With the connection between different interaction models and gate unitaries established, the presented gates are calibrated as follows. (i) First, a frequency sweep using a large-amplitude microwave pulse is performed to exclude spectrally crowded regions. (ii) The pulses are applied at the appropriate frequencies, with p$_1$ and p$_2$ facilitating the transverse coupling, while p$_1$ and p$_3$ inducing the longitudinal coupling. (iii) A complete swap angle $\Theta_\text{XY}=\pi/2$ and zero $\mathrm{ZZ}$ angle, $\Phi_\text{ZZ}=0$, (cf. Fig.~\ref{fig2}) correspond to an $\mathrm{iSWAP}$ gate, up to local $\mathrm{Z}$ gates. The measurement can be performed with the intended gate repeated for an odd number of times, amplifying coherent errors, to find the correct pulse parameters. (iv) With p$_2$ off, there is no transverse coupling, and p$_3$ can be tuned to get $\Phi_\text{ZZ}=\pi/2$, giving the $\mathrm{CZ}$ gate. (v) With p$_1$ and p$_2$ calibrated to give a complete population swap, p$_3$ can be tuned to get $\Phi_\text{ZZ}=\pi/2$, giving the $\mathrm{SWAP}$ gate. (vi) The local phase gates are implemented in software by tracking the frame of the qubit and imparting relative phases on subsequent single-qubit pulses. Their angles are calibrated via least square optimization using tomographic measurement of the final states. Using these calibration steps, we realized a 230-ns-long $\mathrm{iSWAP}$ gate with $\{A_1,A_2,A_3\}/2\pi=\{71.7, 65.2,21.35\}~\mathrm{MHz}$, a 180-ns-long $\mathrm{CZ}$ gate with $\{A_1,A_3\}/2\pi=\{71.2, 28.5\}~\mathrm{MHz}$, and a 260-ns-long $\mathrm{SWAP}$ gate with $\{A_1,A_2,A_3\}/2\pi=\{96.9, 28.5, 14.2\}~\mathrm{MHz}$. As Q$_3$'s $|1\rangle \leftrightarrow |2\rangle$ frequency is close to that of Q$_1$'s $|0\rangle \leftrightarrow |1\rangle$, we implemented the CZ gate by applying p$_1$ and p$_3$ at a frequency 40 MHz blue-detuned from Q$_1$ to avoid any spectator leakage error. \\

\noindent\textbf{Dynamical phase and Z gates} 

    \begin{figure}[t]
        \includegraphics[width=0.49\textwidth]{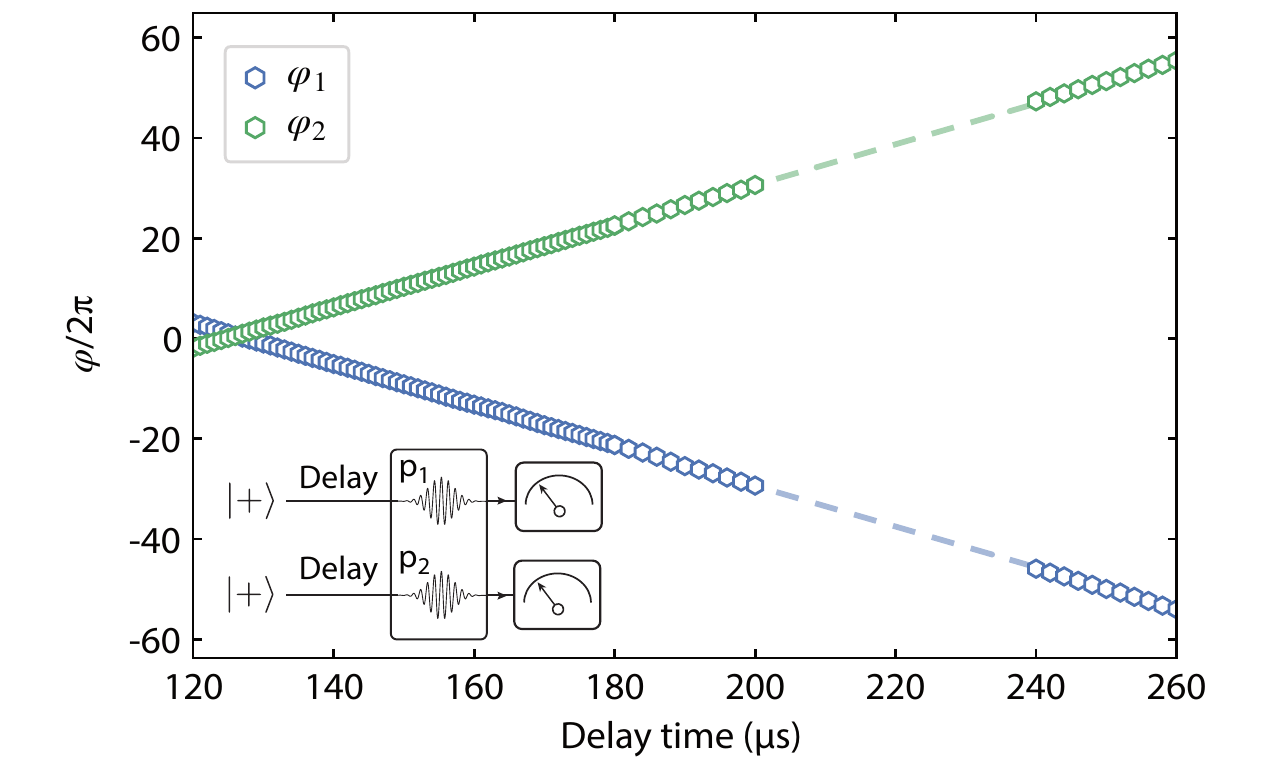}
        \caption{\label{figs5}\textbf{Dynamical phase accumulation}. After preparing the qubits in the superposition state $(|0\rangle+|1\rangle)\otimes(|0\rangle+|1\rangle)/2$, calibrated pulses are employed to facilitate $|10\rangle$ and $|01\rangle$ coherent exchange (inset). Tomographic measurement reveals the qubits' $\mathrm{Z}$ phases which are linearly dependent on the time at which the swap pulses are applied.}
    \end{figure} 

    \noindent In general, integrating a gate resulting from the transverse coupling into a quantum circuit must overcome two technical requirements. First, while the XY interaction is facilitated between the Floquet qubits in the dressed frame, the measurements and single-qubit gates are performed on the bare qubits with different transition frequencies. This leads to an accumulation of single-qubit phases that depend on the time at which the gate is applied, and the accumulation rate is equal to the difference in qubit frequencies \cite{abrams2020implementation}. To verify this effect, we apply pulses p$_1$ and p$_2$ calibrated to implement a perfect swap, then measure the local $\mathrm{Z}$ phases $\varphi_{1,2}$ of the qubits. Upon sweeping the delay time before applying the pulses, we find these phases to increase linearly, and the extracted slope is, within measurement uncertainty, the same as the qubits' frequency difference, as shown in Extended Data Fig.~\ref{figs5}. Including other gates before this transverse swap has the same effect (gap between the data points). Thus, the $\mathrm{Z}$ gate corrections must take this dynamics into account.
    
    Second, the $\mathrm{ZXZXZ}$ single-qubit gate decomposition using virtual-$\mathrm{Z}$ gates comes with the following caveats. This scheme realizes arbitrary single-qubit rotation via Euler decomposition \cite{mckay2017efficient},
    \begin{eqnarray}
        &&\hat{U}(\theta,\phi,\lambda) = \hat{\text{R}}_\text{Z}(\phi) \hat{\text{R}}_\text{X}(\theta) \hat{\text{R}}_\text{Z}(\lambda) \\
        &&= \hat{\text{R}}_\text{Z}\!\left(\phi-\frac{\pi}{2}\right) \hat{\text{R}}_\text{X}\!\left(\frac{\pi}{2}\right) \hat{\text{R}}_\text{Z}(\pi-\theta)\hat{\text{R}}_\text{X}\!\left(\frac{\pi}{2}\right)\hat{\text{R}}_\text{Z}\!\left(\lambda-\frac{\pi}{2}\right).\nonumber
    \end{eqnarray}
    Here, the $\mathrm{Z}$ rotation is implemented virtually by keeping track of all the physical single-qubit $\mathrm{X}$ gates and redefining the rotation axes for all the gates following an intended $\mathrm{Z}$ gate, a procedure known as \textit{phase carrying}. A compatible two-qubit unitary must therefore be a \textit{phase carrier}. This procedure thus breaks down when there is an energy exchange gate in the circuit, for instance, $\mathrm{\sqrt{iSWAP}}$. In addition, since $\hat{X}\hat{X}$ and $\hat{Y}\hat{Y}$ do not commute with single-qubit $\hat{Z}$, it is problematic to implement virtual-$\mathrm{Z}$ gates in the circuit to correct for the additional single-qubit $\mathrm{Z}$ phases induced during the operations~\cite{chen2021compiling}.

    In this work, we utilize the special property of the swap angle $\Theta_\mathrm{XY}=\pi/2$,  $(\hat{\text{R}}_\text{Z}(\varphi_1)\otimes\hat{\text{R}}_\text{Z}(\varphi_2))\times\hat{U}_\mathrm{XXZ}(\pi/2,\Phi_\text{ZZ}) =\hat{U}_\mathrm{XXZ}(\pi/2,\Phi_\text{ZZ})\times (\hat{\text{R}}_\text{Z}(\varphi_2)\otimes\hat{\text{R}}_\text{Z}(\varphi_1))$, to compile the quantum circuits. The induced $\mathrm{Z}$ phases occured during the gate operations can also be corrected by applying virtual-$\mathrm{Z}$ gates afterward~\cite{abrams2020implementation}. The circuit compilation thus requires two additional steps at each $\mathrm{iSWAP}$ or $\mathrm{SWAP}$ cycle, which can be satisfied by computing and adding virtual-$\mathrm{Z}$ gates after them: (i) correction gates for $\mathrm{Z}$ phases must account for the dynamical phase accumulation due to the frame difference, and (ii) the phase tracking for virtual-$\mathrm{Z}$ gates must switch the frames. In practice, condition (i) only requires accounting for the time interval between consecutive gates, and condition (ii) is fulfilled by adding/subtracting the carrying phase, which is an internal part of the compilation software. We note that SU(2) compilations that are compatible with any two-qubit gate have been recently introduced~\cite{chen2021compiling}, so other 
    gates in the $\mathrm{XY}$ family are in principle fully compatible with any platform supporting native virtual-$\mathrm{Z}$ gates.
    \newline

\noindent\textbf{CCZ gate calibration}
    
    \noindent Three different gates must be calibrated for the $\mathrm{CCZ}$ sequence: (i) a $|11\rangle_\text{c} \leftrightarrow |02\rangle_\text{c}$ $\mathrm{iSWAP}$ gate between Q$_2$ and Q$_3$ that does not induce any spectator $\mathrm{ZZ}$ error between Q$_1$ and Q$_2$, (ii) a $\mathrm{CZ}$ gate between Q$_1$ and Q$_2$ that does not induce any spectator error, as this can lead to an effective ZZ entanglement after the shelving for control state $|11\rangle_\text{c}$, and (iii) a $\mathrm{CPhase}$ gate at the end to ensure an overall identity operation on Q$_2$ and Q$_3$, which must not induce extraneous $\mathrm{ZZ}$ between Q$_1$ and Q$_2$. We tune up these gates as follows.
    
    Similar to $\mathrm{XY}$ gates between the computational levels, a frequency sweep is performed to find an appropriate detuning from Q$_3$'s $|1\rangle \leftrightarrow |2\rangle$ transition. A single drive tone is then applied to Q$_3$ at this frequency, its duration and amplitude are varied until a good $|11\rangle_\text{c} \leftrightarrow |02\rangle_\text{c}$ chevron pattern is observed. The $\mathrm{ZZ}$ angle between Q$_1$ and Q$_2$ is measured at the end of the pulse and then cancelled by applying an additional pulse on Q$_1$. It is important to completely cancel this spurious coupling, since its presence results in a non-zero $\mathrm{ZZ}$ phase between Q$_1$ and Q$_2$ for the control state $|11\rangle_\text{c}$, rendering the shelving scheme ineffective. We found the best detuning to be $22~\mathrm{MHz}$ below Q$_3$'s $|1\rangle \leftrightarrow |2\rangle$ transition, and optimal pulse duration to be $280~\mathrm{ns}$ with a ramping time of $120~\mathrm{ns}$. Similarly, a pulse on Q$_3$ is added during the CZ gate operation on Q$_1$ and Q$_2$ to null any spurious effect on the shelved state. To satisfy the more stringent requirements, we increase the ramping time to 90 ns and gate time to 210 ns.
    
    After applying three pulses consecutively, the residual conditional phase $\Phi_\mathrm{ZZ}$ between Q$_2$ and Q$_3$ is measured. We then apply a $\mathrm{CPhase}$ gate between them to invert this conditional phase, and at the same time cancel the residual ZZ coupling between Q$_1$ and Q$_2$. To satisfy the requirements, the gate consists of three microwave pulses applied to all three qubits. The pulses on Q$_2$ and Q$_3$ are tuned up first to negate all the entanglement between them. Then, the remaining $\mathrm{ZZ}$ phase between Q$_1$ and Q$_2$ is cancelled by applying a small-amplitude pulse to Q$_1$. The tuned up CPhase gate is 180-ns long, with a ramp time of 80-ns, bringing the total sequence duration to 950-ns. The calibration is performed manually in this simple fashion, so we expect even better gate performance with future optimization.
    \\
\newline
\noindent\textbf{Cycle benchmarking}
     \begin{figure}[t]
        \includegraphics[width=0.5\textwidth]{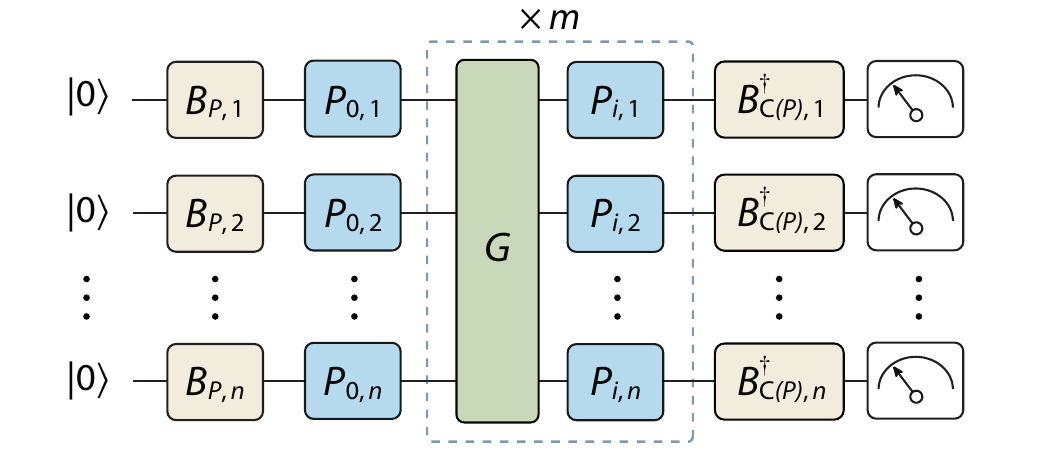}
        \caption{\label{figscb}\textbf{Cycle benchmarking sequence}. The sequence consists of a cycle of preparation gates $B_P$, followed by a cycle of Pauli gate $P$. The gate of interest $G$ is interleaved between cycles of randomly-sampled Pauli gates for $m$ different iterations. Finally, the qubits are rotated to the original eigenstates via a cycle of $B^\dagger_{C(P)}$'s. A reference sequence has the identity gate replacing the $G$ cycle.}
    \end{figure}   
    
    \noindent Cycle benchmarking (CB) \cite{erhard2019characterizing} is a scalable benchmarking protocol for characterizing errors and noise in cycles containing parallel gate operations. Unlike randomized benchmarking (RB), which tailors all errors into a global depolarizing channel via Clifford twirling, CB tailors all errors into stochastic Pauli channels via Pauli twirling. To measure the errors for a Pauli channel $P \in \{I, X, Y, Z \}^{\otimes n}$ for a cycle of operations containing $n$ qubits, the register of qubits is first prepared in an eigenstate of $P$ via single-qubit basis operations $B_P$, followed by a cycle of randomly-sampled Pauli gates $P$, after which the cycle or gate of interest $G$ is interleaved between alternating cycles of randomly-sampled Pauli gates for $m$ iterations, and finally the register of qubits is rotated back to the eigenstate of $P$ via refocusing gates $B^\dagger_{C(P)}$. Each Pauli channel $P$ is fitted to an exponential decay function $A_P f^m_P$ for a circuit depth $m$, where the fidelity of each Pauli channel,
    \begin{figure*}[t]
        \includegraphics[width=\textwidth]{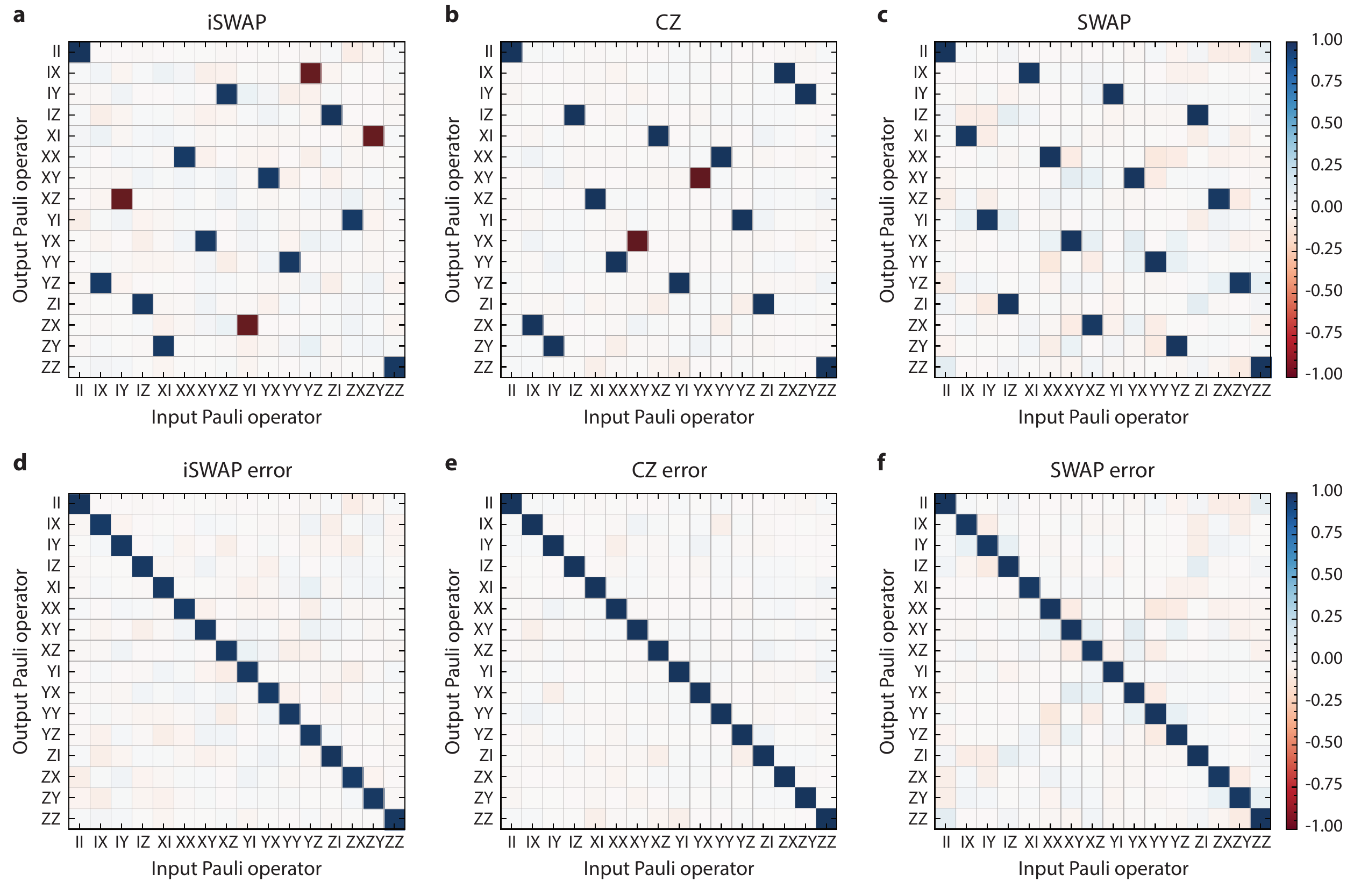}
        \caption{\label{figs4}\textbf{Quantum process tomography results}. Experimentally obtained PTM's corresponding to \textbf{a,} $\mathrm{iSWAP}$, \textbf{b,} $\mathrm{CZ}$, and \textbf{c,} $\mathrm{SWAP}$ gates. Their ideal counterparts are used to compute the respective error PTMs (panels \textbf{d}-\textbf{f}).}
    \end{figure*}
    \begin{equation}
        f_P = \Tr [C(P)^\dagger \Tilde{C}(\rho)],
    \end{equation}
    is captured by the overlap of the results of the ideal circuit $C(P)$ with the noisy implementation $\Tilde{C}(\rho)$, where $\rho$ represents the initial state of the $n$-qubit system in a +1-eigenstate of $P$; the constant $A_P$ represents the state preparation and measurement (SPAM) error. By measuring the performance of the interleaved gate cycle $G$ at two different circuit depths $\{m_1, m_2\}$ for $K$ different Pauli channels, one may estimate the process fidelity via
    \begin{equation}
        \mathcal{F}_\text{p} = \frac{1}{K} \sum_{P \in \mathbb{P}} \left( \frac{\sum^L_{l=1}f_{P,m_2,l}}{\sum^L_{l=1}f_{P,m_1,l}} \right)^\frac{1}{m_2 - m_1},
    \end{equation}
    where $\sum^L_{l=1}$ represents the sum over the $L$ different randomizations for each Pauli channel $P$, and $\tfrac{1}{K} \sum_{P \in \mathbb{P}}$ represents the average over all measured Pauli channels. Here, $\mathbb{P}$ is a subset of the $n$-qubit Pauli group that has been sampled from the full set of $4^n$ possible channels. The number of different Pauli channels $K = |\mathbb{P}| \le 4^n$ sets the precision of the fidelity estimate~\cite{erhard2019characterizing}. The process infidelity is given as $e_\text{p} = 1 - \mathcal{F}_\text{p}$.

    Much like interleaved RB (IRB), which measures the fidelity of a dressed gate composed of the interleaved gate $G$ with a cycle of random Clifford gates, CB measures the fidelity of a dressed cycle composed of the interleaved cycle $G$ and a cycle of random Pauli gates. Similar to IRB, one can estimate the average gate infidelity of just the interleaved cycle $G$ by taking the ratio of the process fidelities of the dressed $D$ and reference $R$ cycles,
    \begin{equation}\label{eq:cb_eG}
        e_\text{g} = \frac{d - 1}{d} \bigg(1 - \frac{\mathcal{F}_\text{p}^{D}}{\mathcal{F}_\text{p}^{R}} \bigg),
    \end{equation}
    where $d=2^n$ is the dimension of the Hilbert space for $n$ qubits. It should be noted that estimating the isolated gate fidelity via IRB or CB can be subjected to large systematic bounds, with upper and lower bounds that may differ by orders of magnitude depending on the unitarity of the gate or cycle~\cite{carignan2019bounding}. However, it has been shown that CB tightens the upper- and lower-bounds on the fidelity estimate relative to IRB~\cite{mitchell2021hardware}, due to the fact that random Pauli gates are more efficiently decomposed into native operations than random Clifford gates. We note that since the conditions of the qubits, including coherence times, naturally fluctuate, while the cycles of interest are not measured simultaneously, Eq.~(\ref{eq:cb_eG}) should be viewed as an estimation.\\
 
    \begin{figure*}[t]
        \includegraphics[width=\textwidth]{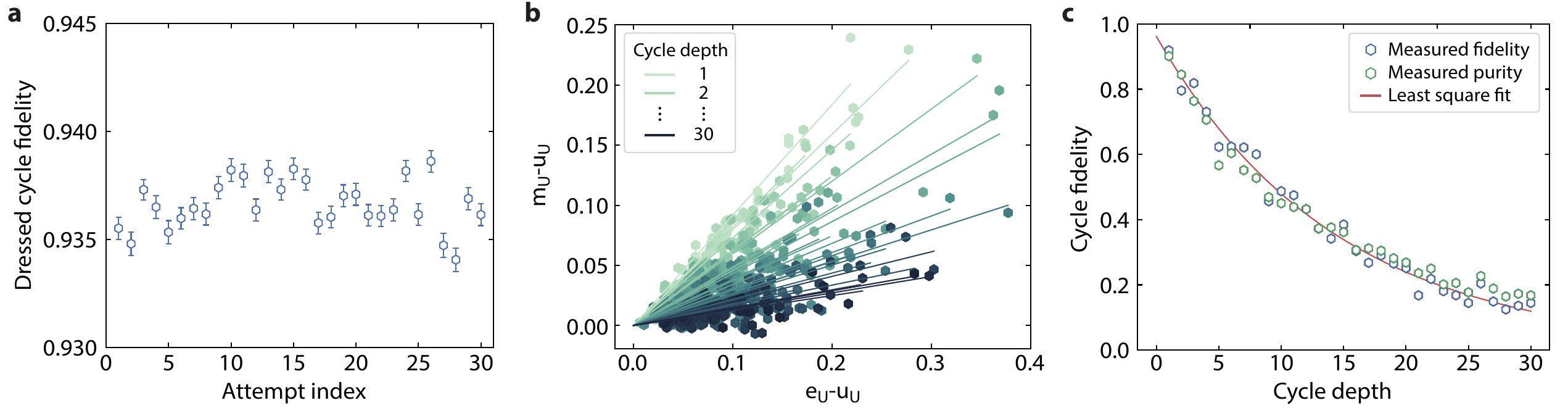}
        \caption{\label{figs8}\textbf{Extended verification of CCZ}. \textbf{a,} Fluctuation of the dressed process fidelity given by CB over a time span of eight hours. \textbf{b,} XEB circuit fidelity at different cycle depths. \textbf{c,} The overall decrease of the average XEB circuit fidelity at larger depths reveals the estimated dressed process fidelity of $93.3(2)\%$. Note that these fidelities include the errors from single-qubit gates used to construct random cycles. We additionally extract the speckle purity at each cycle depth, which is close to the cycle fidelity.}
    \end{figure*}
\noindent\textbf{Cross-Entropy Benchmarking}

    \noindent Cross Entropy Benchmarking (XEB) \cite{arute} is an additional SPAM-free method we leverage for studying the errors and noise associated with the three-qubit CCZ gate. XEB benefits from being able to benchmark non-Clifford unitaries and was instrumental in the first experimental demonstration of quantum advantage. In the XEB protocol, the errors of a multi-qubit gate are tailored into a global depolarizing channel via interleaved cycles of randomly chosen local $SU(2)$ gates. Given sufficient tailoring of the noise, one expects to observe scrambling behavior that manifests as the distribution of probabilities $p$ for observing a particular bitstring following the Porter-Thomas distribution $P(p) = (d-1)(1-p)^{d-2}$, where $d=2^n$ is the dimension of the measured Hilbert space. For a sufficiently randomized circuit, the circuit error can be conveniently thought of as the deviation of the measured bitstring distributions from a uniform distribution.  
    
    We seek to make this relationship precise. We denote all possible bitstrings as $x_i$ for $i = 1,...,2^n$ where $n$ is the number of qubits that the unitary of interest act on. We assign $q(x_i)$ as the measured distribution, and define the linear cross-entropy between two probability distributions $p_1(x)$ and $p_2(x)$ as
    \begin{equation}
        H(p_1,p_2) = \sum_x p_1(x)p_2(x), 
    \end{equation}
    where the sum runs over the full support of the probability distributions. The XEB circuit fidelity is then given as
    \begin{equation}
        \mathcal{F}_\mathrm{XEB} = \frac{H(p,q) - H(p,u)}{H(p,p) - H(p,u)}\equiv \frac{m_U-u_U}{e_U-u_U},
    \end{equation}
    where $u(x)=1/d$ is the uniform probability distribution on the bitstrings. This expression can be understood as the difference in the ideal to measured and ideal to uniform cross entropies, normalized by the difference if the measured distribution was to perfectly match the ideal distribution (i.e. $p(x_i) =  q(x_i)$ for all $i$). 
    
    An additional feature of the XEB routine is that at each depth $m$ in the XEB circuits, the so-called speckle purity $\gamma$ can be used to estimate the decoherence-limited cycle fidelity. It is calculated using the measured bitstring probability distributions as
    \begin{equation}
        \gamma(m) = \text{Var}(p_m)\frac{d^2(d+1)}{(d-1)}.
    \end{equation}
    Here $p_m$ is the measured probability distribution for bitstrings at depth $m$ in our XEB circuits, where $\mathrm{Var}(p_m)=(d-1)/d^2(d+1)$ for ideal pure states.
    \\

\noindent\textbf{Extended gate verification}

    \noindent \emph{Two-qubit gates:} The implemented unitaries are also benchmarked using quantum process tomography (QPT), which can be used to approximate the residual coherent errors. For example, single-qubit $\mathrm{Z}$ gates can be approximated using the extracted Pauli transfer matrix (PTM) $\mathcal{R}$ using Nelder-Mead least square optimization. The PTMs for the realized gates are shown in Extended Data Fig.~\ref{figs4}\textbf{a}-\textbf{c}. They can be reverted using the ideal PTMs to find the error PTMs $\mathcal{R}_\mathrm{error} = \mathcal{R}_\mathrm{exp} \mathcal{R}_\mathrm{ideal}^{-1}$, which is useful in prescribing the residual errors to Pauli channels (Extended Data Fig.~\ref{figs4}\textbf{d}-\textbf{f}). The process fidelity $\mathcal{F}_\text{p}$ and gate fidelity $\mathcal{F}_\text{g}$ are extracted following $\mathcal{F}_\text{p}=\mathrm{Tr}(\mathcal{R}_\mathrm{ideal}^T\mathcal{R}_\mathrm{exp})/d^2$ and $\mathcal{F}_\text{g}=(d\mathcal{F}_\text{p} + 1)/(d+1)$, where $d=4$ is the dimension of the two-qubit system. For the displayed $\mathrm{iSWAP}$, $\mathrm{CZ}$, and $\mathrm{SWAP}$ PTMs, the extracted gate fidelities are $98.4\%$, $99.8\%$, and $98.9\%$, respectively. Since the extracted PTMs are sensitive to SPAM errors and maximum-likelihood estimation, the fidelities should only be viewed as rough approximation, and their best utilization is for calibration.

    \emph{CCZ gate:} To gauge the stability of the calibration, we continuously monitor the dressed cycle fidelity over 30 consecutive iterations taking over eight hours using CB (Extended Data Fig.~\ref{figs8}\textbf{a}). Although the sequence consists of multiple gates, we only observe small variation in the fidelity which do not break the limits imposed by coherence time fluctuations. We additionally verify the gate fidelity using cross-entropy benchmarking (XEB), achieving an average XEB fidelity of $93.3(2)\%$ (Extended Data Fig.~\ref{figs8}\textbf{b,c}), consistent with the CB result. Note that this fidelity includes the error from single-qubit gates used to construct random cycles. XEB also allows us to extract the speckle purity, which is shown to be approximately the same as the cycle fidelity (Extended Data Fig.~\ref{figs8}\textbf{c}). This implies that the gate is coherence-limited. \newline

\begin{figure*}[t]
        \includegraphics[width=\textwidth]{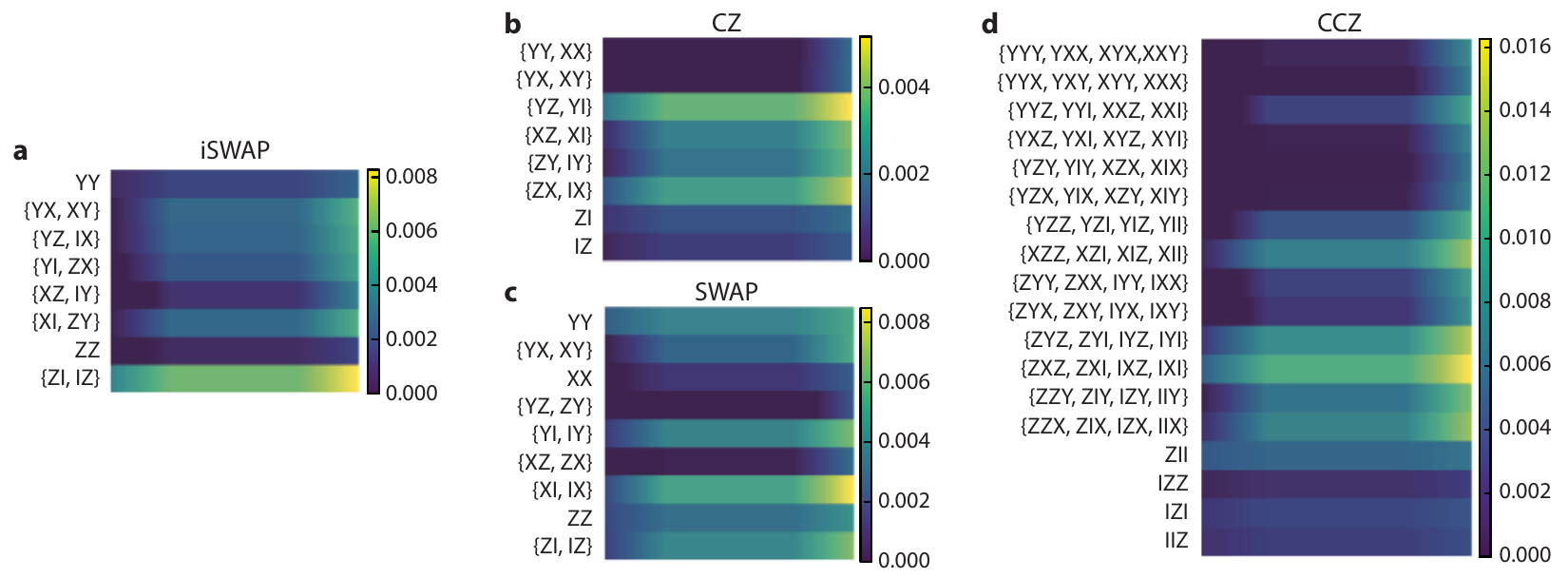}
        \caption{\label{fig:knr}\textbf{Error classification}. Pauli
noise reconstruction (PNR) results showing the dominant errors impacting \textbf{a,} $\mathrm{iSWAP}$, \textbf{b,} $\mathrm{CZ}$, \textbf{c,} $\mathrm{SWAP}$, and \textbf{d,} $\mathrm{CCZ}$ gates. The y-axes label the most dominant Pauli Kraus errors for each gate, the cell color denotes the error rate, and the cell gradient defines the 95\% confidence interval uncertainty. Pauli errors grouped in curly brackets denote degenerate gate errors which cannot be distinguished due to the effect of the gate itself.}
\end{figure*}
    
\noindent\textbf{Error budgets} 
   
    \noindent To estimate the dephasing-limited fidelity, we first assume that the qubit decays at a rate $\Gamma_1$ and dephases at a rate $\Gamma_\phi$, where these rates are related to the relaxation time $T_1$ and decoherence time $T_2$ as $\Gamma_1 = 1/T_1$, $\Gamma_2=1/T_2$. A Pauli transfer matrix (PTM) of a single-qubit decoherence channel for a duration $\tau$ is given as
    \begin{equation}
        \mathcal{E}(\tau) = 
        \begin{pmatrix}
        1 & 0 & 0 & 0 \\
        0 & e^{-\Gamma_2\tau} & 0 & 0 \\
        0 & 0 & e^{-\Gamma_2\tau} & 0 \\
        0 & 0 & 0 & e^{-\Gamma_1\tau} 
        \end{pmatrix}.
    \end{equation}
    
    In the absence of non-Markovian errors such as leakage and crosstalk, the PTM of a Pauli-twirled $n$-qubit channel is simply given by the tensor product $\mathcal{E}^{\otimes n}$. The process fidelity limited by decoherence can thus be written as
    \begin{equation}\label{eqn:decoherence_fidelity}
    \begin{split}
        \mathcal{F}_\text{p} &= \frac{1}{4^n}\mathrm{Tr}[\mathcal{E}^{\otimes n}] \\
        & = \frac{1}{4^n} \prod_{i=1}^n\left(1+e^{-\Gamma_1^{(i)} \tau} + 2e^{-\Gamma_2^{(i)} \tau} \right),
    \end{split}
    \end{equation}
    where $\Gamma_1^{(i)}$ and $\Gamma_2^{(i)}$ denote the energy relaxation and dephasing rates of qubit $i$, respectively. 
    
    Using Eq.~(\ref{eqn:decoherence_fidelity}), we estimate the coherence-limited fidelities using the best undriven $T_1$ and $T_2^E$ times for the upper bound, and the worst dynamical values for the lower bound. The lower fidelity bounds for $\mathrm{iSWAP}$, $\mathrm{CZ}$, $\mathrm{SWAP}$, and $\mathrm{CCZ}$ gates are $\{99.2\%,~99.4\%,~99.1\%,~95.1\%\}$ and  the upper bounds are $\{99.6\%,~99.7\%,~99.5\%,~97.1\%\}$. Although this approach only gives approximate limits, they are already remarkably close to the obtained fidelities. We note that the $\text{CCZ}$ sequence contains four different pulses involving different qubit pairs, so it is difficult to choose the appropriate coherence times for the analysis. However, the speckle purity measurement (Extended Fig.~\ref{figs8}\textbf{c}) reveals that it should be dephasing-limited.
    
    In practice, the ramp times ($\sim$50-100 ns) constitute a majority of the pulses used for the gates ($\sim$200 ns), as opposed to the long pulses with mostly fixed amplitudes used in the dynamical coherence measurements (Fig.~\ref{fig3}\textbf{b}). Besides, different gates require different combinations of pulse amplitudes and ramp times, making it challenging to systematically select the right coherence values for our approximation. Therefore, we believe that it is reasonable to estimate the fidelity limits using the presented approach. A more rigorous estimation may ideally involve a numerical Floquet simulation with the integration of a time-dependent mapping. Future research to elucidate such dynamics, shedding light on the dephasing process, will help establish the required techniques to study spurious effects associated with Floquet-engineered interactions.
    

    To reconstruct the dominant error channels impacting the $\mathrm{iSWAP}$, $\mathrm{CZ}$, $\mathrm{SWAP}$, and $\mathrm{CCZ}$ gates, we utilize Pauli noise reconstruction (PNR)~\cite{flammia2020efficient}, also referred to as cycle error reconstruction~\cite{hashim2021randomized}. PNR reconstructs the Pauli Kraus errors affecting each gate from a set of CB measurements (see the Supplemental Material of Ref.~\cite{hashim2021randomized} for a detailed discussion). For example, in Extended Data Fig.~\ref{fig:knr}, we observe that the dominant error channel for the $\mathrm{iSWAP}$ gate is a Z error on either qubit, even though ZZ appears to be largely suppressed (note that IZ and ZI cannot be distinguished, due to the fact that the $\mathrm{iSWAP}$ gate transforms one into the other). For the $\mathrm{CZ}$ gate, we observe that Y-type errors on Q$_1$ or X-type errors on Q$_2$ are the most dominant. The $\mathrm{SWAP}$ gate is mostly affected by X- or Z-type errors on either qubit. While the error generators for the $\mathrm{CCZ}$ are more difficult to interpret, we consistently observe that Z-type errors on Q$_1$, X- or Y-type errors on Q$_2$, and Z-type errors on Q$_3$ are the most dominant.

\begin{figure*}
        \includegraphics[width=\textwidth]{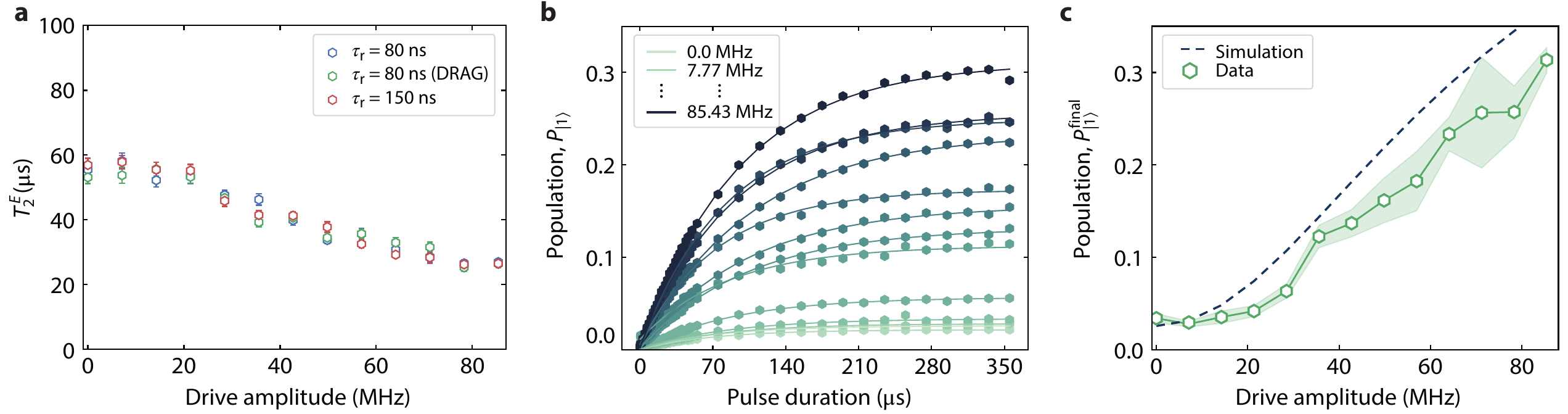}
        \caption{\label{figs7}\textbf{Dressed decoherence}. \textbf{a,} Dependence of Q$_1$'s echo time $T_2^E$ on the drive amplitude and pulse shape. 
        \textbf{b,} Q$_1$'s excited state population after the mapping using a pulse with parametrized amplitude and duration. The qubit is first initialized in $|0\rangle$. \textbf{c,} Comparison between obtained excited state populations after a 355-$\mu$s-long pulse at various driving amplitudes (green markers) and numerically simulated values (blue dashed line).}
    \end{figure*}
\noindent\textbf{Dressed heating and dephasing} \\
\noindent We utilize the single-shot readout capability of the measurement setup (cf.~Extended Data Fig.~\ref{figs2}\textbf{a}) to selectively measure the dynamics of Q$_1$ under the drive. To measure the relaxation process, the qubit is first prepared in state $|1\rangle$ by applying a $\pi$ pulse to a pre-selected $|0\rangle$ state. Then, a microwave pulse with variable amplitude and duration is applied to the qubit. The ramp time is chosen to be 100~ns to ensure adiabaticity for the amplitude values shown. We discard the data in which small oscillation appears, which correspond to the non-adiabatic regime. The qubit is measured dispersively after the off-resonant pulse, and then post-selected in the $|0\rangle$ state. The excitation dynamics is measured in a similar fashion, with the qubit prepared in the ground state and post-selected in the $|1\rangle$ state. For $T_2^E$, a $\pi$ pulse on the undriven qubit is inserted between two Floquet drive pulses with the same ramp and duration times. This refocusing pulse is orthogonal in phase to the $\pi/2$ projection pulses. To ensure the passive reset of the qubit, a delay time of 50~$\mu s$ is added between neighboring sequences.

To verify that the reduction in the obtained $T_2^E$ times is not due to nonadiabatic effects, we repeat the measurements with different ramp times $\tau_\text{r}$ and DRAG coefficients $\lambda_\text{DRAG}$, and still observe a consistent decrease with respect to the pulse amplitude, as shown in Extended Data Fig.~\ref{figs7}\textbf{a}. Intriguingly, we also find a consistent heating effect that increases the excited state population after the mapping, suggesting a change in the effective qubit temperature. Notably, although the population of the qubit increases monotonically with respect to the drive amplitude (Extended Data Fig.~\ref{figs7}\textbf{b}), the extracted relaxation and excitation rates do not show any obvious trend (Fig.~\ref{fig3}\textbf{b}). 

We simulate this effect numerically by projecting the decoherence mechanisms onto the Floquet basis. Using the experimentally obtained ratio $\nu=P^\text{final}_{|1\rangle}/P^\text{final}_{|0\rangle}=0.027$ of the undriven qubit in equilibrium (355 $\mu$s after initialization), we assume a finite relaxation rate $\Gamma_\downarrow= (T_1)^{-1}/(1+\nu)$, excitation rate $\Gamma_\uparrow = (T_1)^{-1}-\Gamma_\downarrow$, and pure dephasing rate $\Gamma_\phi=(T_2^E)^{-1}-(2T_1)^{-1}$, where $T_1$ and $T_2^E$ are taken from Table~\ref{tab:qparameters}. Next, we simply map the jump operators $\hat{L}_-=\sqrt{\Gamma_\downarrow}|0\rangle\langle1|$, $\hat{L}_+=\sqrt{\Gamma_\downarrow}|1\rangle\langle0|$, and $\hat{L}_\phi=\sqrt{\Gamma_\phi/2}\hat{\sigma}_z$ onto the Floquet basis, then evolve the system starting from the ground state using the Lindblad master equation. The result shown in Extended Data Fig.~\ref{figs7}\textbf{c} agrees well with the experimentally obtained data (previously shown in the inset of Fig.~\ref{fig3}\textbf{b}).

\end{document}